\documentclass[11pt]{article}
    \usepackage{amssymb}
    \usepackage{amsmath}
    \usepackage{amstext}
    \usepackage{graphicx,epsfig}
    \usepackage{epsfig}
    \usepackage{verbatim} 
    \usepackage{fancybox}
    \usepackage{color}
    \usepackage{ulem}
    \usepackage{enumitem}
    \usepackage{youngtab}
    \usepackage{subfigure}
    \usepackage{bbm}
    \usepackage{parskip}
    \usepackage{braket}
    \usepackage[numbers,sort&compress]{natbib}
    \usepackage{ytableau}
    \usepackage[utf8]{inputenc}

    \usepackage{tikz}
    \usetikzlibrary{positioning, trees,decorations, decorations.markings, decorations.pathmorphing, arrows, shapes.geometric, snakes, patterns}
    \pgfdeclaredecoration{complete sines}{initial}
    {
    \state{initial}[
    width=+0pt,
    next state=upsine,
    persistent precomputation={\pgfmathsetmacro\matchinglength{
    \pgfdecoratedinputsegmentlength / int(\pgfdecoratedinputsegmentlength/\pgfdecorationsegmentlength)}
    \setlength{\pgfdecorationsegmentlength}{\matchinglength pt}
    }] {}
    \state{upsine}[width=\pgfdecorationsegmentlength,next state=downsine]{
    \pgfpathsine{\pgfpoint{0.25\pgfdecorationsegmentlength}{0.5\pgfdecorationsegmentamplitude}}
    \pgfpathcosine{\pgfpoint{0.25\pgfdecorationsegmentlength}{-0.5\pgfdecorationsegmentamplitude}}
    }
    \state{downsine}[width=\pgfdecorationsegmentlength,next state=upsine]{
    \pgfpathsine{\pgfpoint{0.25\pgfdecorationsegmentlength}{-0.5\pgfdecorationsegmentamplitude}}
    \pgfpathcosine{\pgfpoint{0.25\pgfdecorationsegmentlength}{0.5\pgfdecorationsegmentamplitude}}
    }
    \state{final}{}
    }

    \newcommand{\Comment}[1]{{}}
    \definecolor{darkblue}{rgb}{0.15,0.35,0.55}
    \definecolor{reddish}{rgb}{0.65, 0.2, 0.2}
    \usepackage[linktocpage=true]{hyperref}
    \hypersetup{
    colorlinks=true,
    citecolor=darkblue,
    linkcolor=reddish,
    urlcolor=darkblue,
    pdfauthor={},
    pdftitle={},
    pdfsubject={}
    }

    \definecolor{green3}{RGB}{44, 160, 44}

    \newcommand{\be}{\begin{equation}}
    \newcommand{\ee}{\end{equation}}
    \newcommand{\bea}{\begin{eqnarray}}
    \newcommand{\eea}{\end{eqnarray}}
    \newcommand{\beas}{\begin{eqnarray*}}
    \newcommand{\eeas}{\end{eqnarray*}}
    \newcommand{\nn}{\nonumber}

    \definecolor{darkred}{rgb}{0.7,0.3,0.3}
    \definecolor{darkgreen}{rgb}{0.2,0.7,0.3}
    \definecolor{lightgreen}{rgb}{.816,.94,.753}
    \definecolor{greyish}{rgb}{.8,.8,.8}
    \definecolor{darkblue2}{rgb}{0.3,0.4,0.9}
    \usepackage{pifont}
    \usepackage{xcolor,colortbl}

    \def\({\left(}
    \def\){\right)}

    \newcommand{\pb}{\mathbf{p}}
    \newcommand{\kb}{\mathbf{k}}
    \newcommand{\xb}{\mathbf{x}}

    \newcommand{\la}{\langle}
    \newcommand{\ra}{\rangle}

    \def\gsim{ \lower .75ex \hbox{$\sim$} \llap{\raise .27ex \hbox{$>$}} }
    \def\lsim{ \lower .75ex \hbox{$\sim$} \llap{\raise .27ex \hbox{$<$}} }

    \definecolor{myblue}{rgb}{ 0.188, 0.478,0.858}

    \def\xyma{\xymatrix@M.7em}
    \def\xymas{\xymatrix@M.1em}

    \title{}
    \author{}
    
    \numberwithin{equation}{section}

\begin{document}
    \tikzset{
    photon/.style={decorate, decoration={snake}, draw=magenta},
    graviton/.style={decorate, decoration={snake}, draw=black},
    sgal/.style={decorate , dashed, draw=black},
    scalar/.style={decorate , draw=black},
    mgraviton/.style={decorate, draw=black,
    decoration={coil,amplitude=4.5pt, segment length=7pt}}
    electron/.style={draw=blue, postaction={decorate},
    decoration={markings,mark=at position .55 with {\arrow[draw=blue]{>}}}},
    gluon/.style={decorate, draw=magenta,
    decoration={coil,amplitude=4pt, segment length=5pt}} 
    }
    \renewcommand{\thefootnote}{\fnsymbol{footnote}}
    ~

    \begin{center}
    {\Large \bf On the EFT of Conformal Symmetry Breaking \\
    }
    \end{center} 
    
    \vspace{1truecm}
    \thispagestyle{empty}
    \centerline{\Large Kurt Hinterbichler,${}^{\rm a,}$\footnote{\href{mailto:kurt.hinterbichler@case.edu} {\texttt{kurt.hinterbichler@case.edu}}} Qiuyue Liang,${}^{\rm b,}$\footnote{\href{mailto:qyliang@sas.upenn.edu} {\texttt{qyliang@sas.upenn.edu}}} Mark Trodden,${}^{\rm b,}$\footnote{\href{mailto:trodden@physics.upenn.edu} {\texttt{trodden@physics.upenn.edu}}} }

    \vspace{.5cm}

    \centerline{{\it ${}^{\rm a}$CERCA, Department of Physics,}}
    \centerline{{\it Case Western Reserve University, 10900 Euclid Ave, Cleveland, OH 44106, USA}} 
    \vspace{.25cm}
    
    \centerline{{\it ${}^{\rm b}$Center for Particle Cosmology, Department of Physics and Astronomy,}}
    \centerline{{\it University of Pennsylvania, Philadelphia, Pennsylvania 19104, USA}} 
    
    \vspace{1cm}
    \begin{abstract}
    \noindent
    
Conformal symmetry can be spontaneously broken due to the presence of a defect or other background, which gives a symmetry-breaking vacuum expectation value (VEV) to some scalar operators. We study the effective field theory of fluctuations around these backgrounds, showing that it organizes as an expansion in powers of the inverse of the VEV, and computing some of the leading corrections. We focus on the case of space-like defects in a four-dimensional Lorentzian theory relevant to the pseudo-conformal universe scenario, although the conclusions extend to other kinds of defects and to the breaking of conformal symmetry to Poincar\'e symmetry.
        
    \end{abstract}
    
    
    \setcounter{tocdepth}{2}
    \newpage
    \tableofcontents
    \renewcommand*{\thefootnote}{\arabic{footnote}}
    \setcounter{footnote}{0}

    \newpage
    
    \section{Introduction}
    
  In the real world, no system is infinite, and boundary effects inevitably become important.  Therefore, boundaries and defects in quantum field theory are natural subjects of interest in studying systems with finite extent or where multiple regions or phases are joined by junctions or localized impurities.  In particular, recent years have seen increased activity in the study of such boundaries and defects in the context of conformal field theory (CFT) (see e.g. \cite{Cardy:2004hm,Billo:2016cpy,Andrei:2018die} for overviews.)
    
    A defect in a CFT can be of any dimension.  One of the simplest cases is when the defect is straight.  A straight $d$ dimensional defect breaks the $D$ dimensional conformal group down to the $d$ dimensional conformal group, plus the group of rotations around the defect.  In a Lorentzian CFT, such a defect can be space-like or time-like, but the more frequently-discussed case is that of a time-like defect, where the defect represents a spatial boundary or interface in the system.  
    
    In contrast, an example where space-like defects are relevant is in some non-inflationary early universe scenarios featuring a pre-big bang phase, in which the universe is nearly flat rather than accelerating.  Some well-known examples of this type are the ekpyrotic scenario \cite{Khoury:2001wf,Buchbinder:2007ad} and genesis-type models \cite{Creminelli:2006xe,Creminelli:2010ba}.  
    The general class of such models that makes use of a space-like conformal defect is the pseudo-conformal universe scenario \cite{Rubakov:2009np,Creminelli:2010ba,Libanov:2010nk,Hinterbichler:2011qk,Hinterbichler:2012mv,Hinterbichler:2012fr,Libanov:2012ev,Libanov:2010ci,Libanov:2015iwa}.   
    In these scenarios, it is postulated that the universe before reheating is described by a CFT on a nearly Minkowski spacetime whose conformal algebra is broken spontaneously by a time-dependent vacuum expectation value (VEV) of some dimension $\Delta$ scalar primary operator $\Phi$ taking the form
    \be 
    \la\Phi\ra_{\rm }= {C^\Delta \over {(-t)}^{\Delta}} \ ,
    \label{vevforminte}
    \ee
    where the dimensionless constant $C$ signals the strength of the symmetry breaking.
    The VEV \eqref{vevforminte} breaks the four-dimensional Lorentzian conformal symmetry down to a three-dimensional Euclidean conformal symmetry
    \be 
    \mathfrak{so}(4,2)\rightarrow \mathfrak{so}(4,1) \ .
    \label{symbreakpatternee} 
    \ee
    As $t\rightarrow 0$ from below, the VEV \eqref{vevforminte} diverges and the universe then reheats and transitions to the standard post-big bang radiation domination phase.  The reheating surface at $t=0$ is the space-like defect in the CFT.  This CFT scenario can also be a given a five-dimensional AdS dual description \cite{Hinterbichler:2014tka,Libanov:2014nla,Hinterbichler:2015pta,Libanov:2016gwi,Carrillo-Gonzalez:2020ejs}

    Our primary interest here will be in studying the effective field theory (EFT) that describes fluctuations around the symmetry breaking vacuum described by \eqref{vevforminte}.
    In \cite{Hinterbichler:2012mv} an effective field theory for studying such fluctuations was described.  Here we will further explore some of the systematics of this effective theory.  In particular, we will see how the EFT expansion organizes itself as a power series in various powers of $1/C$, and we will compute some of the leading corrections to the 2-point function. 
    
    Since our original motivation came from studying the pseudo-conformal universe scenario, we will specialize in this paper to the case of a space-like co-dimension 1 defect in a four-dimensional Lorentzian CFT.  However, nothing we do depends crucially on this, and everything will generalize straightforwardly to other dimensions and signatures for both the CFT and the defect.
    
  This EFT can also be used in the simpler case where conformal symmetry is broken to Poincar\'e symmetry via a constant VEV $ \la\Phi\ra_{\rm }\sim f$.  We will see that the EFT naturally organizes as an expansion in various powers of $1/f$.  For example, the two-point function $\la \phi(x)\phi(0)\ra$ organizes as an expansion in various powers of $1/(fx)$, which is good at long distances (complementary to the short distance limit which can be probed by the operator product expansion), and that the leading 1-loop quantum correction is universal, independent of the higher derivative operators in the EFT.

    \textbf{Conventions:} $D$ is the spacetime dimension, and we use the mostly plus metric signature. 
    The curvature conventions are those of \cite{Carroll:2004st}.

    \section{The EFT of conformal symmetry \label{confsymsec}}
    
    The EFT we seek should describe the fluctuations of fields around the symmetry breaking VEV \eqref{vevforminte}.  This is the EFT that describes the spontaneous breaking of conformal symmetry, and which was studied many years ago as a prototype for spontaneously broken spacetime symmetries~\cite{Callan:1970yg,Volkov:1973vd}.  It is equivalent~\cite{Bellucci:2002ji,Elvang:2012st,Creminelli:2013ygt} to the theory of a co-dimension 1 brane fluctuating in a fixed background anti-de Sitter space.  The same EFT also plays a key role in the proof of the $a$-theorem ~\cite{Komargodski:2011vj,Komargodski:2011xv}, and its $S$-matrix satisfies non-trivial soft theorems \cite{Boels:2015pta,Huang:2015sla,DiVecchia:2015jaq,DiVecchia:2017uqn}.

    One well-known way to construct this EFT is from the coset perspective (see e.g. \cite{Goon:2012dy,Hinterbichler:2012mv}).  However, in the following we describe an alternative and more direct method of constructing it for arbitrary conformal weights, which will prove useful in the rest of the paper, and which starts with fields that linearly realize conformal symmetry.
    
    \subsection{Direct construction}
    
    Our goal is to construct the EFT of a weight $\Delta$ scalar conformal primary field $\Phi$. The symmetries that must be maintained are the usual linearly-realized conformal symmetries,
    \bea 
    \label{scalesme} 
    \delta \Phi &=&- ( x^\mu\partial_\mu+\Delta) \Phi \ ,\\ 
    \delta_\mu\Phi &= &-\( 2x_\mu x^\nu\partial_\nu -x^2\partial_\mu+2x_\mu\Delta\)\Phi \, ,  
    \label{conformalsyme}
    \eea
    which are the scale transformation and special conformal transformations, respectively.  
    
    We construct the EFT by writing all conformally invariant terms, order by order in powers of derivatives.  
    We allow for terms which are non-analytic in the fields, because we will ultimately be expanding around a conformally non-invariant VEV\footnote{This is similar to the philosophy of the ``Higgs EFT" as opposed to the ``Standard Model EFT" in the context of electroweak symmetry breaking \cite{Falkowski:2019tft,Cohen:2020xca}.}.
    
    Scale invariance is easy to impose; it is equivalent to demanding that each term in the Lagrangian density has total operator dimension equal to the spacetime dimension $D$, so that there are no dimensionful couplings.  We will assume throughout that $\Delta\not=0$ and $D>2$, since other subtleties arise otherwise.
    
   At zeroth order in derivatives, the only scale invariant term is
    \be 
    {\cal L}_0=\Phi^{D\over \Delta} \ .
    \label{lagdeeq0}
    \ee
    This term is also invariant under special conformal transformations, so this is our complete zeroth order Lagrangian.
    
    At second order in derivatives, the only possible scale invariant term, up to total derivatives, is
    \be 
    {\cal L}_2=\Phi^{{D-2\over \Delta}}{(\partial \Phi)^2\over\Phi^2} \ . 
    \label{lagdeeq2}
    \ee
    This is also invariant under the special conformal transformations, so this is our 2-derivative Lagrangian.
    
    At fourth order in derivatives, there are three possible scale invariant terms, up to total derivatives: 
    \be 
    {\Phi^{{D-4\over \Delta}} {(\square\Phi)^2\over \Phi^2} },\ \ \ { \Phi^{{D-4\over \Delta}}{(\partial\Phi)^2\square\Phi \over\Phi^3} },\ \ \ { \Phi^{{D-4\over \Delta}}{(\partial\Phi)^4\over \Phi^4} } \ .
    \ee   
    However, imposing special conformal invariance, only two linear combinations of these three terms are invariant.  For later convenience we choose these combinations to be
    \bea 
    && {\cal L}_4={\Phi^{{D-4\over \Delta}}} \left[  {(\square\Phi)^2\over \Phi^2}   -\frac{(2 \Delta-D +2) (2 \Delta-D +4)}{4 \Delta ^2}  {(\partial\Phi)^4\over\Phi^4} \right],\ \ \  \nn\\
    &&{\cal L}_4'=\Phi^{{D-4\over \Delta}}\left[ {(\partial\Phi)^2\square\Phi  \over\Phi^3}-\frac{2 \Delta-D +3}{2 \Delta }{(\partial\Phi)^4\over\Phi^4}\right] \ . 
    \label{lagdeeq4}
    \eea
    This construction can be continued to all higher orders in derivatives; at each derivative order there will be some finite number of independent scale invariant terms up to total derivatives, some subspace of these will be fully conformal invariant, and a basis of this subspace forms the EFT Lagrangian at this derivative order.  
    
    The full Lagrangian is then the sum of all these terms, with arbitrary coefficients, organized as a derivative expansion,
    \be 
    {\cal L}=c_0 {\cal L}_0+c_2{\cal L}_2+c_4 {\cal L}_4+c_4'{\cal L}_4'+\cdots \ .
    \label{lagrangiantote}
    \ee
    We will let $\{c\}_4\equiv \{c_4,c_4'\}$, $\{c\}_6 \equiv \{c_6,c_6',c_6'',\ldots\}$, etc. denote the sets of coefficients of terms at each derivative order.
    
    \subsection{Comparison to the coset construction}
    
    The usual approach to constructing this theory is the coset approach, equivalent to the geometric method as described in \cite{Goon:2012dy,Hinterbichler:2012mv}.  Here we will extend this approach to arbitrary $\Delta$ and see that it is equivalent to the above direct approach.  
    
    We consider the conformally flat metric
    \be 
    g_{\mu\nu} = \left(\Phi^{1/\Delta}\over \Lambda\right)^2\eta_{\mu\nu}\ ,
    \label{conforflatmet}
    \ee
where $\Lambda$ is some artificially introduced energy scale that allows us to ensure that the metric has the correct dimensions. This quantity is purely a calculational convenience, and will cancel out at the end of our calculations.  
    We then write diffeomorphism invariants using this metric, with an overall scale set by $\Lambda$.  These diffeomorphism invariants are equivalent to the basic building blocks of the coset construction.
    The set of all diffeomorphism invariants at a given derivative order will in general be larger than the number of conformally invariant Lagrangians, because some invariants may be degenerate on the specific conformally flat metric \eqref{conforflatmet}.  But no invariant Lagrangians will be missed in this way, with one exception in even $D$ at $D$-th order in derivatives: the Wess-Zumino term (see below.)  
    
    Starting at zeroth order in derivatives, the only diffeomorphism invariant is the cosmological constant.  When evaluated on \eqref{conforflatmet} this reproduces the zeroth order Lagrangian \eqref{lagdeeq0},
    \be 
    \Lambda^D \int d^Dx \ \sqrt{-g} =\int d^Dx \ \Phi^{D\over \Delta}   \, . 
    \ee
    
    At second order, the only diffeomorphism invariant is the Einstein-Hilbert term.  When evaluated on \eqref{conforflatmet} this reproduces the second order Lagrangian \eqref{lagdeeq2} up to total derivatives,
    \be 
    \Lambda^{D-2} \int d^Dx \ \sqrt{-g} R = \frac{(D-2) (D-1)}{\Delta ^2} \int d^Dx \ \Phi^{{D-2\over \Delta}}{(\partial \Phi)^2\over\Phi^2} \, . 
    \ee
    
    At fourth order, there are 3 different curvature invariants up to total derivatives: $R^2$, $R_{\mu\nu}^2$ and $R_{\mu\nu\rho\sigma}^2$.  However we need not consider anything made from the Riemann tensor, since the Weyl tensor vanishes when evaluated on the conformally flat metric \eqref{conforflatmet}.  Thus, we have two possible invariants, and evaluated on \eqref{conforflatmet} they recover, up to total derivatives, linear combinations of the invariants \eqref{lagdeeq4},
    \bea  
    && \Lambda^{D-4} \int d^Dx \ \sqrt{-g} R^2 = \int d^Dx \  \frac{4 (D-1)^2}{\Delta ^2}{\cal L}_4+\frac{4 (D-1)^2 (D-2 \Delta -2)}{\Delta ^3}{\cal L}_4' \,,\nn \\
     && \Lambda^{D-4} \int d^Dx \ \sqrt{-g} R_{\mu\nu}^2 = \int d^Dx \  \frac{(D-1) D}{\Delta ^2} {\cal L}_4+\frac{(D-2) \left(3 D^2-8 D+8\right)-4 (D-1) D \Delta }{2 \Delta ^3}{\cal L}_4' \,.  \nn  \\  
     \label{R2equationtolage2}
\eea
    
    The $2\times 2$ matrix mapping these two curvature invariants to the two Lagrangians has full rank for $D\not=4$, and so in this case we can recover both ${\cal L}_4$ and ${\cal L}_4'$ from the curvatures by solving \eqref{R2equationtolage2} for ${\cal L}_4$, ${\cal L}_4'$.  For $D=4$ however, this matrix has rank 1, and we cannot recover both Lagrangians from the curvature invariants.  In this case both curvature invariants give the same linear combination:
    \be    
    \int d^4x \ \sqrt{-g} R^2 =  \int d^4x \ \sqrt{-g} R_{\mu\nu}^2={12\over \Delta^2} \int d^4x \ {\cal L}_4-{2(\Delta-1)\over \Delta}{\cal L}_4'\,,\ \ \ D=4 \, . 
    \ee
    The other linear combination is a Wess-Zumino term, which cannot be constructed from the curvature invariants (see \cite{Goon:2012dy} for details on how to construct it within the coset formalism\footnote{A quick way to do it is to form the Gauss-Bonnet combination in general $D$,
    \be 
    \Lambda^{D-4} \int d^Dx \sqrt{-g} \left( R_{\mu\nu\rho\sigma}^2-4R_{\mu\nu}^2+R^2  \right)=-\frac{2 (D-4) (D-3) (D-2)}{\Delta ^3}  \int d^Dx \ {\cal L}_4' \, .
    \ee
    This results in a quantity that is proportional to $D-4$ because in $D=4$ the Gauss-Bonnet combination becomes a total derivative. Stripping off the $D-4$ factor yields the Wess-Zumino term \cite{Nicolis:2008in}.  
    }).   In any even $D$, a similar Wess-Zumino term exists at $D$-derivative order.

     \section{Breaking to Poincar\'e\label{Poincaresection}}
    
    To start with, we demonstrate the power of this EFT technique by applying it to simplest case, where conformal symmetry breaks to Poincar\'e symmetry.  This will serve as a warmup for our main case of interest, defect CFT, and illustrates most of the general features.
    
    To describe the breaking to Poincar\'e we expand around a constant VEV,
    \be 
    \la \Phi \ra=f^\Delta,\ \ \ \Phi=f^\Delta\left(1+{1\over f^{{D\over 2}-1}}\phi\right) \, , 
    \label{flatfeveve}
    \ee
    where $f$ is a constant with mass dimension 1, and the coefficient for the fluctuation field $\phi$ is chosen so that $\phi$ will be canonically normalized.
    
    We now expand the Lagrangian \eqref{lagrangiantote} in powers of $\phi$. Demanding the absence of the tadpole so that the constant VEV is a solution to the equations of motion requires
    \be 
    c_0=0 \ .
    \ee
    At second order in $\phi$ we then have
    \be  {\cal L}_{\phi,2}=-{1\over 2}(\partial\phi)^2+{c_4\over f^2}(\square\phi)^2+{\cal O}\left({\partial^6 \phi^2 \over f^4},\{c\}_6\right)+{\cal O}\left({\partial^8\phi^2 \over f^6},\{c\}_8\right)+\cdots \ , 
    \label{l2flatlag2}
    \ee
    where we have used the freedom to scale the field (and to flip the overall sign of the Lagrangian if necessary) to choose $c_1=-{1\over 2}$ so that $\phi$ has a canonically normalized 2-derivative kinetic term.
    These quadratic terms organize as a power series in $1/f^2$, with coefficients of the higher order terms coming from the higher derivative terms in \eqref{lagrangiantote}.
    At cubic order we have
    \be  
    {\cal L}_{\phi,3}={1\over f^{{D\over 2}-1}}\left[{2\Delta-D+2\over 2\Delta}\phi(\partial\phi)^2+{\cal O}\left({\partial^4 \phi^3 \over f^2},\{c\}_4\right)+\cdots\right] \ .
    \ee
    The cubic terms organize as a power series in $1/f^2$, starting at order $1/ f^{{D\over 2}-1}$, with coefficients of the higher order terms coming from the higher derivative terms in \eqref{lagrangiantote}.
    At quartic order we have
    \be  
    {\cal L}_{\phi,4}={1\over f^{{D}-2}}\left[-\frac{(2 \Delta -D+2) (3 \Delta -D+2)}{4 \Delta ^2} \phi^2(\partial\phi)^2+{\cal O}\left({\partial^4 \phi^4 \over f^2},\{c\}_4\right)+\cdots\right] \ .
    \ee
    The quartic terms organize as a power series in $1/f^2$, starting at order $1/ f^{{D}-2}$, with coefficients of the higher orders coming from the higher derivative terms in \eqref{lagrangiantote}.  This pattern continues: the terms at $n$-th order organize in powers of $1/f^2$, starting at order $1/ f^{(n-2)\left ({D\over 2}-1\right)}$.
    
    \subsection{2-point function}
    
  We can use this effective theory to compute correlators systematically as a power series expansion in $1/f$.   These will depend on the free coefficients $c_4,c_4',\cdots$, parametrizing the higher derivative terms in the action, which also serve as counterterms to absorb divergences.  As we will see, there are also some universal parts that do not depend on these coefficients.
    
   To illustrate, consider the two point function $\la \Phi(x)\Phi(0)\ra$.  The broken conformal symmetry puts no constraint on the functional form of this correlator beyond the usual  constraints from Poincar\'e invariance that tell us it must be a function of the magnitude of the invariant distance between the two points.  Expanding using \eqref{flatfeveve} and using the assumption $\la \phi\ra=0$, we have
    \be 
    \la \Phi(x)\Phi(0)\ra=f^{2\Delta} \left[1+ {1\over f^{D-2}} \la \phi(x)\phi(0)\ra\right] \ . 
    \label{PhiPhiexpee}\ee
    Now $\la \phi(x)\phi(0)\ra$ can be computed using Feynman diagrams in the effective theory.  
    
    From power counting \cite{Burgess:2007pt,Goon:2016ihr} we can see that in dimensional regularization the diagrams contributing to the momentum space 2-point function will scale as
    \be 
    \la\phi(p)\phi(-p)\ra \sim {1\over p^2}\left(p\over f\right)^{(D-2)L}\left(c \,{p\over f}\right)^{\sum_{n,k}(k-2)V_{n,k}} \, ,
    \label{pestimatee}
    \ee
    where $V_{n,k}$ is the number of vertices with $n$ fields and $k$ derivatives in the diagram, $L$ is the number of loops, and $c$ stands generically for the coefficients of the higher-derivatve terms.  
    
    Suppose that we are only interested in the correlator away from $x^2=0$, so that we can ignore terms analytic in $p^2$ (which only contribute when $x^2=0$). 
    
    At tree level, $L=0$, the only corrections come from the higher-order vertices in \eqref{l2flatlag2}, and these contributions are all analytic in $p^2$.  So all that remains at separated points is the zeroth order propagator of the free kinetic term,
    \be 
    \la\phi(p)\phi(-p)\ra_{0-{\rm loop}}=-{i\over p^2} \ .
    \label{freepropagatore}
    \ee
    
\begin{figure}[h!]
    \begin{center}
    \epsfig{file=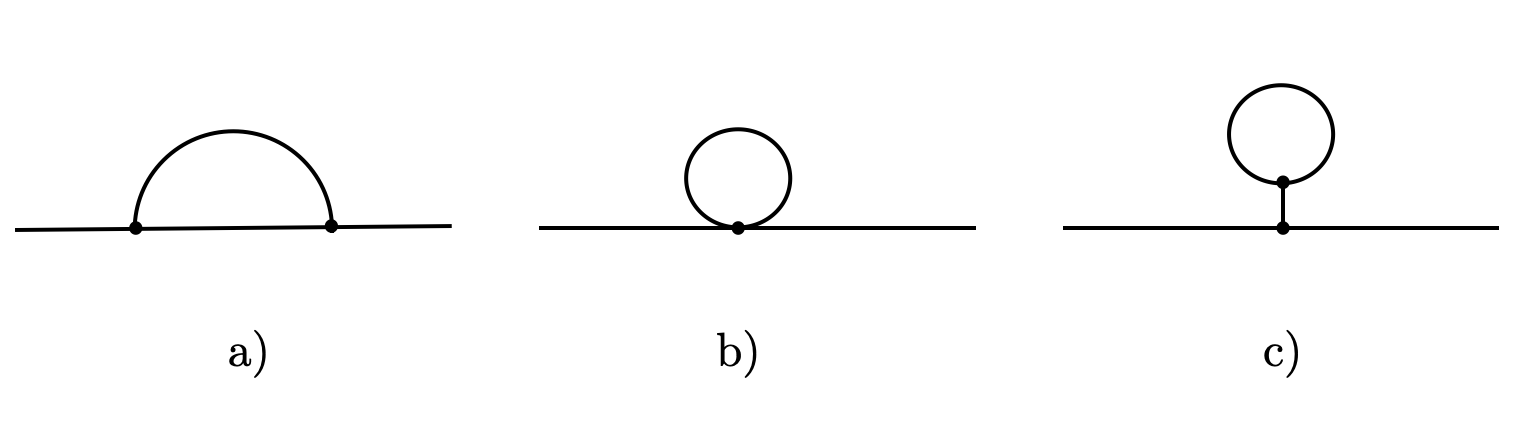,height=1.9 in,width=6in}
    \caption{Contributions to the 2-point function at 1-loop.}
    \label{figure1}
    \end{center}
    \end{figure}

    The leading corrections at separated points come from $L=1$, with no insertions of higher-order vertices.  The diagrams are shown in Figure \ref{figure1}.
    
    The only diagram that contributes is the first one, since the rest vanish in dimensional regularization.  For $D$ even, there is a divergence that is proportional to
    \be 
    \sim {1\over \epsilon} {(2\Delta-D+2)^2\over \Delta^2} {1\over f^{D-2}} (p^2)^{{D\over 2}-2} \, ,
    \ee
    where $\epsilon$ is the dimensional regularization parameter.  This is analytic in $p^2$, is absorbed by one of the higher derivative tree-level counterterms, and does not contribute at separated points.  The finite part is logarithmic in $p^2$,
    \be 
    \la\phi(p)\phi(-p)\ra_{1-{\rm loop}}= {i^{D+1}\over 2^{2D-1}\pi^{D-1\over 2}\Gamma\left({D\over 2}-1\right)} {(2\Delta-D+2)^2\over \Delta^2}  {1\over f^{D-2}} (p^2)^{{D\over 2}-2}\log(p^2/\mu^2) \, , \ \ \ D\ {\rm even} \ ,
    \ee
    where $\mu$ is the dimensional regularization scale whose value is adjusted through the same counterterm that absorbs the divergence, and does not affect separated points.
    
    For odd $D$, the $L=1$ contribution \eqref{pestimatee} involves an odd power of $p^2$, so it is not analytic.  This means that there cannot be any UV divergence, because the counterterms can only absorb analytic terms.  Indeed, the loop is finite in dimensional regularization and gives
    \be 
    \la\phi(p)\phi(-p)\ra_{1-{\rm loop}}= {i^D\over 2^{2D-1}\pi^{D-3\over 2}\Gamma\left({D\over 2}-1\right)} {(2\Delta-D+2)^2\over \Delta^2}  {1\over f^{D-2}} (p^2)^{{D\over 2}-2} \, , \ \ \ D\ {\rm odd} \ .
    \ee
   
    Going back to position space, the 1-loop contribution gives a contribution to the correlator at separated points that is proportional to $\sim {1/ x^{2(D-2)}}$.  Including the position space version of the tree level part \eqref{freepropagatore}, which is proportional to $\sim{1/ x^{D-2}}$, and substituting back into \eqref{PhiPhiexpee}, we see that the full 2-point function takes the schematic form
    \bea 
    \la \Phi(x)\Phi(0)\ra&=&f^{2\Delta} \left[1+ {1\over f^{D-2}}\left( \la \phi(x)\phi(0)\ra_{0-{\rm loop}} +\la \phi(x)\phi(0)\ra_{1-{\rm loop}}+\cdots\right)  \right] \nn\\
    &\sim &  f^{2\Delta} \left[1+ {1\over f^{D-2}} {1\over x^{D-2}}  + {1\over f^{2(D-2)}}{1\over x^{2(D-2)}}+\cdots \right]  \, ,
    \eea
where we have ignored a possible dimensionless constant in each term.    

   The loop expansion is an expansion in powers of ${1/ ( fx )^{D-2}}$, and the one-loop contribution is independent of the higher-derivative coefficients and is a universal prediction of the EFT.  At higher orders in the expansion, insertions of the higher derivative terms into loops will bring in further factors of $1/ ( fx )^2$.   The EFT expansion is thus valid for large $x\gg 1/f$, which is good at long distances/low energies, as expected.  Note that this is opposite to the regime $x\ll 1/f$ for which the operator product expansion of the underlying CFT can be used to expand the correlator in powers of $xf$ \cite{Karananas:2017zrg}.

    \section{Defect CFT breaking}
    
   We now turn to the main case of interest, the breaking of conformal symmetry to a lower dimensional conformal symmetry in the presence of a maximally symmetric defect.  We expand the EFT around the desired time-dependent symmetry breaking VEV \eqref{vevforminte},
    \be 
    \la\Phi\ra_{\rm }= {C^\Delta\over {(-t)}^{\Delta}} \ ,
    \label{vevforminte2}
    \ee
    where the dimensionless constant $C$ parametrizes the strength of the symmetry breaking, and we take $C>0$ without loss of generality. 
    
    A general way to parametrize the fluctuations around this VEV is to write
    \be 
    \Phi= \la\Phi\ra F(\phi) \ ,
    \ee
    with $\phi$ the field parametrizing the fluctuations, and $F$ some arbitrary function that is analytic at the origin and satisfies $F(0)=1$.
    The metric \eqref{conforflatmet} in term of $\phi$ is then
    \be 
    g_{\mu\nu} =   \left(\Phi^{1/\Delta}\over \Lambda\right)^2\eta_{\mu\nu}={1\over \Lambda^2} {C^2\over t^2} F(\phi)^{2\Delta} \eta_{\mu\nu}=  F(\phi)^{2\Delta} \bar g_{\mu\nu} \ ,
    \ee
    where $\bar g_{\mu\nu}$ is a fake (meaning we introduce it as an unphysical mathematically useful artifact) de Sitter metric with a Hubble scale,
    \be 
    \bar g_{\mu\nu}\equiv {1\over H^2 t^2}\eta_{\mu\nu},\ \ \ H\equiv {\Lambda\over C} \ .
    \ee
    Thus, the effective theory of fluctuations is obtained by expanding curvature invariants around de Sitter space, and the unbroken $\mathfrak{so}(D,1)$ symmetry of the full conformal symmetry $\mathfrak{so}(D,2)$ is manifest.
    
    In \cite{Hinterbichler:2012mv}, an exponential parametrization $F(\pi)=e^{\pi}$ was used.
    For our purposes, it will be more straightforward to instead use the linear parametrization $F(\phi)=1+{ (-t)^{{D\over 2}-1} \over C^{{D\over 2}-1}}\,\phi$, so that we have
    \be 
    \Phi={C^\Delta\over {(-t)}^{\Delta}} \left(1+{ (-t)^{{D\over 2}-1} \over C^{{D\over 2}-1}}\,\phi\right) \, ,
    \label{symmbreakexpe} 
    \ee 
    where we have chosen the factor in front of the fluctuation $\phi$ to ensure that this field will be canonically normalized when we expand the action.
    
    We now expand the Lagrangian \eqref{lagrangiantote} in powers of $\phi$ using \eqref{symmbreakexpe}.   At linear order there is a tadpole term. Demanding that this vanishes up to total derivatives, so that the background \eqref{vevforminte2} is a solution to the equations of motion, gives the condition\footnote{We will show later that the one-loop correction does not contribute to the tadpole term.}
    \be 
    c_0 + {1\over C^2} c_2 \Delta^2 +{1\over C^4}\left[ c_4  {\Delta^2\over 4}\left(3 D^2-10 D+4-4 \Delta ( D-1)\right) -c_4'{\Delta^3(D-1)\over 2} \right]+{\cal O}\left({1\over C^6},\{c\}_6\right)+\cdots=0 \, . 
    \ee
    This condition is an expansion in powers of $1/C^2$, with coefficients arising from the Lagrangian terms with higher derivatives.   We use this condition to solve for $c_0$, keeping only up to the order in $1/C^2$ that we are ultimately interested in.  In addition, we will use the freedom to rescale the field $\Phi$ (and to flip the overall sign of the Lagrangian if necessary) to choose 
    \be 
    c_2=-{1\over 2} \, ,
    \ee
    which will ensure that the kinetic term for the fluctuation $\phi$, as defined in \eqref{symmbreakexpe}, is canonically normalized.
    
   At higher orders we get the following terms up to total derivatives: at second order in $\phi$ we have
    \bea  
    {\cal L}_{\phi,2}&=& {1\over 2}\dot\phi^2-{1\over 2}(\vec\nabla \phi)^2+\frac{D(D+2)}{8 t^2} \phi^2 \nn\\
    &&+{1\over C^2}\bigg[ c_4\bigg(  t^2 \ddot \phi^2-2 t^2 (\vec\nabla\dot\phi)^2+(2 \Delta  (D-1)-D^2+5 D-2) \dot\phi^2+ t^2 (\vec\nabla^2\phi)^2   \nn\\
    &&  -\left(  2(D-1)\Delta-(D-1)(D-4) \right) (\vec\nabla\phi)^2  +\frac{  8D(D-1)(D+2)\Delta-{D(D+2)(5D^2-18D+8)  }}{16 t^2}\phi^2 \bigg)  \nn\\ 
     && + c_4' \Delta (D-1)\bigg( \dot\phi^2-(\vec\nabla\phi)^2 +\frac{D (D+2)}{4 t^2}\phi^2\bigg) \bigg]\nn\\
    &&+{\cal O}\left({1\over C^4},\{c\}_6\right)+\cdots \ .
    \label{quadraticcCactionee}
    \eea
    These quadratic terms organize as a power series in $1/C^2$, with coefficients coming from the higher order terms in \eqref{lagrangiantote}.
    
    At cubic order in $\phi$ we have
    \bea  
    {\cal L}_{\phi,3}= {1\over C^{{D\over 2}-1}}  {(-t)^{{D\over 2}-1}\over 2\Delta} \bigg[ &&  (2 \Delta -D+2)\left(\phi \dot\phi^2-\phi(\vec\nabla\phi)^2\right)+{2 \left(D^2+6 D-4\right) \Delta-D^3-6 D^2+12 D-8 \over 12 t^2}\phi^3 \nn\\
    &&+{\cal O}\left({1\over C^2},\{c\}_4\right)+\cdots \bigg] \, . \label{cubicvertle3e}
    \eea 
    These terms also organize as a power series in $1/C^2$, starting at order ${1/ C^{{D\over 2}-1}} $.
    
    At quartic order in $\phi$ we have
    \bea  
    \label{eq,quarticL}
    {\cal L}_{\phi,4}= {1\over C^{D-2}}  {(-t)^{D-2}\over 4\Delta^2} \bigg[ &&  (6 \Delta ^2-5 \Delta  (D-2)+(D-2)^2)\left(\phi^2 \dot\phi^2-\phi^2(\vec\nabla\phi)^2\right) \nn\\
    &&+{ 4 (10 D-9) \Delta ^2-3 \left(13 D^2-28 D+20\right) \Delta +9 D^3-30 D^2+44 D-24  \over 12 t^2}\phi^4 \nn\\
    &&+{\cal O}\left({1\over C^2},\{c\}_4\right)+\cdots \bigg] \, .
    \eea 
    These terms again organize as a power series in $1/C^2$, starting at order ${1/ C^{D-2}} $.  This pattern continues: the terms at $n$-th order in the fluctuation organize in powers of $1/C^2$, starting at order $1/ C^{(n-2)\left ({D\over 2}-1\right)}$.

    \section{Defect CFT 2-point function}
    
    We can now use the EFT to compute correlators of $\Phi$ in the broken vacuum \eqref{vevforminte2} as an expansion in powers of $1/C$.  Consider for example the 2-point function $\la \Phi(t,\xb )\Phi(t',\xb')\ra$.  The broken conformal symmetry dictates that the general form of this correlator is an arbitrary function $F$ of the cross ratio $\xi$ \cite{Fubini:1976jm,Cardy:1984bb,McAvity:1993ue,McAvity:1995zd,Liendo:2012hy,Billo:2016cpy,Lauria:2017wav,Sarkar:2018eyy,Carrillo-Gonzalez:2020ejs},
    \be 
    \label{eq,crossratio}
    \la \Phi(t,\xb )\Phi(t',\xb')\ra = {F(\xi)\over (-t)^\Delta (-t')^\Delta},\ \ \ \xi\equiv { -(t-t^\prime)^2+(\xb-\xb')^2\over 4 (-t)(-t')} \, . 
    \ee
    Substituting in the expansion \eqref{symmbreakexpe} and using that $\la \phi\ra=0$,  we have
    \be  
    \la \Phi(t,\xb )\Phi(t',\xb')\ra= {C^{2\Delta}\over (-t)^\Delta(-t')^\Delta } \left[ 1+{1\over C^{D-2}}{ (-t)^{{D\over 2}-1} (-t')^{{D\over 2}-1}}  \la \phi(t,\xb )\phi(t',\xb')\ra \right] \, . 
    \ee
    We can then compute the fluctuation correlator $\la \phi(t,\xb )\phi(t',\xb')\ra$ from the expanded Lagrangian, which will take the form of a loop expansion in powers of $1/C^{D-2}$ and higher derivative corrections of order $1/C^2$. 
    
    We now turn to computing some of the leading corrections.  The tree level correlator is easily evaluated.  The next to leading order correction is more difficult and we will restrict ourselves to computing the equal time, late time limit, $t = t^\prime  \to 0$.
    
    \subsection{Leading part}
    
    We compute the leading $C$-independent part of the fluctuation correlator $\la \phi(t,\xb )\phi(t',\xb')\ra$ from the $C$ independent part of the quadratic action; the first line of \eqref{quadraticcCactionee}.   This has the same form as that of a massive scalar on de-Sitter space with mass to Hubble ratio $m^2/H^2=-D$.\footnote{Note that this is the mass value for a $k=1$ Galileon/DBI shift-symmetric scalar in de Sitter \cite{Bonifacio:2018zex,Bonifacio:2021mrf,Goon:2011qf,Goon:2011uw,Burrage:2011bt}.  The shift symmetry is the broken conformal symmetry.} 
    
    Expanding the field in mode functions
    \be 
    \phi(t,\xb)={1\over (2\pi)^{(D-1) /2}} \int d^{D-1}  \kb \left[\ a_\kb f_k (t) e^{i\kb\cdot \xb}+a_\kb ^\dag f_k^\ast (t) e^{-i\kb\cdot \xb}\right] \ ,
    \ee 
   where the mode functions should satisfy the equation
    \be  
    f_k(t)''+\left[k^2-{D(D+2)\over 4t^2}\right]f_k(t)=0 \ ,
    \label{mukhanovee}
    \ee
    which depends only on the magnitude $k\equiv |\kb|$ of the spatial momentum.
We will assume the standard Bunch-Davies boundary conditions
    \be 
    f_k(t)\underset{t\rightarrow -\infty }{\longrightarrow} {1\over \sqrt{2k}}e^{-ik t} \ .
    \label{BDcondition}
    \ee
    which gives the solution
    \be 
    \label{eq,modefunction}
    f_k(t)={\sqrt{\pi}\over 2}\sqrt{-t} \, e^{i\pi\left( D+2\over 4 \right)}H_{D+1\over 2}^{(1)}\left(-kt\right) \ .
    \ee
    
Using the usual creation/annihilation operator relations, $ [a_\kb,a^\dag_{\kb'}]=\delta^{D-1}(\kb-\kb')$, $[a_\kb,a_{\kb'}]=[a_\kb^\dag,a^\dag_{\kb'}]=0$, $a_\kb|0\ra=\la 0|a_\kb^\dag=0$, we can now compute the spatial Fourier transform of the 2-point function,
    \bea 
    &\la \phi(t,\kb)\phi(t',\kb')\ra = \int d^{D-1}\xb d^{D-1} \xb' e^{i\kb \cdot \xb}e^{i\kb'\cdot \xb'}\la \phi(t,\xb)\phi(t',\xb')\ra= (2\pi)^{D-1}\delta^{D-1}(\kb+\kb')f_k(t)f_{k'}^\ast(t') \ . & \nn\\
    \eea

    \subsection{Next to leading order corrections}
 We now compute the 2-point correlation function beyond leading order. We will specialize to the case $D=4$, but will keep $D$ explicit in the following expressions in order to make dimensional regularization easier to implement.  The contributions to this part are the tree level diagrams with higher derivative terms from the quadratic action in \eqref{quadraticcCactionee}, and the loop corrections from the cubic interaction and quartic interaction as shown in figure \ref{figure1}. 
    
    We will make use of the Mellin-Barnes representation \cite{Sleight:2019mgd} to compute the corrections via the in-in formalism and to regulate the loop \cite{Premkumar:2021mlz}. 
    The mode function \eqref{eq,modefunction} expressed in the Mellin-Barnes representation has the following form,
    \begin{equation}
        f_{k}(t) = \frac{e^{i \pi \nu-\frac{i \pi}{4}}}{\sqrt{4 \pi}} \int_{c-i\infty}^{c+i\infty} \frac{d s}{2 \pi i} \Gamma\left(s+\frac{ \nu}{2}\right) \Gamma\left(s-\frac{ \nu}{2}\right)\left(-\frac{i}{2} k \right)^{-2 s}(-t)^{-2 s+\frac{1}{2}} \,, \label{MBrepmodefunctionsee}
    \end{equation}
    where $\nu \equiv \frac{D+1}{2}$, and the contour is chosen to be a straight line parallel to the imaginary axis, intersecting the real axis at $c$, and is closed to the left.  We require $c > \left|\mathfrak{Re} \frac{\nu}{2} \right|$ to include all the poles within the contour. In flat spacetime, the Fourier transform provides a convenient representation for mode functions since its variable $k^0$ is the energy, and the resulting $\delta$-functions that arise during calculations enforce conservation of energy because of time-translations symmetry. Similarly, in de Sitter space, the Mellin-Barnes representation is convenient because the variable $s$ can be thought of as the eigenvalue of the dilatation symmetry, with the corresponding $\delta$-functions enforcing its conservation.

 The equal time in-in 2-point function evaluated at time $t_0$ is given by
      \begin{equation}
        B_{2} \equiv \lim_{t \to -\infty} \braket{0|~\mathcal{U}^\dagger (t_0,t)  \phi(t_0, \mathbf{k})  \phi(t_0, \mathbf{k}'   )\mathcal{U} (t_0,t) |0}\,,
    \end{equation} 
    where $\mathcal{U}$ is the full time evolution operator, given in terms of the interaction picture Hamiltonian $\mathcal{H}_{\text{int}}(t)$ via,
    \begin{equation}
        \mathcal{U} = 1- i \int_{-\infty}^{t_0} dt \mathcal{H}_{\text{int}}(t) -\frac{1}{2} \int_{-\infty}^{t_0} dt_1 \int_{-\infty}^{t_0} dt_2 \mathcal{T}\left\{\mathcal{H}_{\text{int}} (t_1 )\mathcal{H}_{\text{int}} (t_2) \right\}+ \cdots \, , \label{uintexpe}
    \end{equation}
   where $\mathcal{T} \{\cdots\}$ represents time-ordering.

    Define the bulk-to-boundary propagator for a time-ordered vertex as:
    \begin{equation}
        G_{B+}(k,t) = f_k(t) f^*_{k}(t_0) \,,
    \end{equation}
    and the bulk-to-bulk propagator between two time-ordered vertices as, 
    \begin{equation}
        G_{++} (k, t_1,t_2) = \theta(t_1 - t_2)  f^*_k(t_1) f_k(t_2)  + \theta(t_2 - t_1) f^*_k(t_2) f_k(t_1) \,.
    \end{equation}
    The propagators for anti-time ordered vertices can then be obtained through conjugation,
    \begin{equation}
         G_{B-}(k,t) =  G_{B+}(k,t)^* = f^*_k(t)  f_k(t_0 ) \,,
    \end{equation}
    \begin{equation}
        G_{--} (k,t_1, t_2) = G_{++}(k,t_1,t_2)^* = \theta(t_1 - t_2) f^*_k(t_2)f_k(t_1)  + \theta(t_2 - t_1) f^*_k(t_1)f_k(t_2)  \,.
    \end{equation}
    The bulk-to-bulk propagators connecting a time-ordered vertex at time $t_1$ and an anti-time ordered vertex at time $t_2$ are given by
    \begin{equation}
    \label{eq,pmpropagator}
        G_{+-}(k , t_1 ,t_2) =f^*_k(t_2)f_k(t_1) ~, \quad G_{-+}(k, t_1 ,t_2) = G_{+-}(k, t_2 ,t_1) =  f_k(t_2) f^*_k(t_1)\,.
    \end{equation}
   
    With these ingredients at hand, we can start computing the $1/C^2$ corrections from the various diagrams.  We will see that, analogously to the Poincar\'e case in Section \ref{Poincaresection}, only the diagram on the left in Figure \ref{figure1} contributes non-trivially.  
   
    \subsubsection{Higher derivative term}
    
    We start with the tree level correction with a single vertex drawn from the terms with power $1/C^2$ in \ref{quadraticcCactionee}. Once the equations of motion for the external lines are taken into account, the term proportional to $c_{4^\prime}$ reduces to a boundary term,
    \begin{equation}
        -i \frac{c_{4^\prime} \Delta (D - 1)}{C^2 } f_k(t_0)\dot{f}_k(t_0)f_k^\ast(t_0) f_k^\ast(t_0)\,.
    \end{equation}
    After taking the late time limit and summing with the anti-time ordered vertex, this boundary term contributes as
    \begin{equation}
    \label{eq,c4prime}
       - \frac{3 c_{4^\prime} \Delta }{2 C^2 } \frac{1}{(k^5 t_0^4)}\left(9+ 3 (-k t_0)^2 +   (-k t_0)^4 \right) = -  \frac{3 c_{4^\prime} \Delta }{ C^2 } f_k(t_0) f_k^\ast(t_0)\,,
    \end{equation}
    where we have explicitly set $D= 4$. In fact, it is true for any even dimension $D$ that $ \text{Re}\left( - i f_k(t_0)\dot{f}_k(t_0) f_k^\ast(t_0)^2 \right)= -\frac{1}{2} f_k(t_0) f_k^\ast(t_0) $ in the late time limit. In the same spirit, we can integrate by parts in the $c_4$ term, organizing the action as
    \begin{eqnarray}
     S_2 &=&  \int \frac{dt  d^{D-1} \kb }{(2\pi )^{D-1}} \left[ \left(2\Delta (D-1)  -D^2 + 5D -2 \right) \left(\dot \phi^2 - k^2 \phi^2 + \frac{D(D+2)}{4t^2} \phi^2  \right)\right. \nonumber\\
     &+& \left. 2 k^2 \phi^2 - \frac{D^2 (D+2)^2}{16 t^2} \phi^2 + t^2 \ddot{\phi}^2 -2 t^2 k^2 \dot{\phi}^2 + t^2 k^4 \phi^2 \right]\,.
    \end{eqnarray}
  Since the mode functions satisfy the Hankel equation $ \ddot{f}_k (t) = -\frac{4k^2 t^2 - D(D+2)}{4 t^2} f_k(t)$, we can simplify the second line to yield 
    \begin{eqnarray}
      -2 t^2 k^2 \dot{\phi}^2 + 2t^2 k^4 \phi^2 - \frac{1}{2} k^2 D(D+2) \phi^2 + 2k^2 \phi^2 \ ,
    \end{eqnarray}
    which is also a pure boundary term $\partial_t \left(-2 t^2 k^2 \dot \phi \phi + 2 k^2 t \phi^2 \right)$. The extra factor of $k^2 t^2$ makes the boundary terms higher order in the momentum $k$, in comparison with \eqref{eq,c4prime},  
    \begin{eqnarray}
    && - i \frac{c_4}{C^2}\left(\left[(2\Delta (D-1) -D^2 + 5D -2) -2 t_0^2 k^2 \right] f_k(t_0)\dot{f}_k(t_0)f_k^\ast(t_0) f_k^\ast(t_0) + 2 t_0 k^2 f_k(t_0) f_k(t_0)f_k^\ast(t_0) f_k^\ast(t_0) \right) \nonumber\\
    &  &\xrightarrow{\lim_{t_0 \to 0}}     - i \frac{c_4}{  C^2} (2 \Delta (D-1) -D^2 + 5D -2)  f_k(t_0) \dot{f}_k(t_0) f_k^\ast(t_0)^2 \,.
    \end{eqnarray}

    \subsubsection{Quartic loop}
    We next turn to computing the correction at $\mathcal{O} (1/C^2)$ from the loop with a single quartic vertex, as shown in part b) of figure \ref{figure1}.

    After integrations by parts and use of the lowest order equations of motion on the external lines, the quartic lagrangian \eqref{eq,quarticL}  takes the form  
    \begin{equation}
        S_4 \propto \int dt \frac{d^{D-1} \kb}{(2\pi)^{D-1}} (-t)^{D-4} \phi^4\, ,
    \end{equation}
and the contribution of this to the 2-point correlation function takes the corresponding form
    \begin{eqnarray}
    && B_{2q}^{1+}=  - \frac{i}{2} e^{\frac{\pi i (2D+1 )}{2}}f^\ast_k(t_0)^2 \int_{-\infty}^{t_0} f_k(t)^2 \int \frac{d^{D-1}\pb}{(2\pi)^{D-1} } f_p(t)f_p^\ast(t)\,.
    \end{eqnarray} 
    Inserting the Mellin-Barnes representation of the mode function \eqref{eq,modefunction}, we have 
    \begin{eqnarray} 
    - \frac{i e^{\frac{\pi i (2D+1 )}{2}} }{2}  f^\ast_k(t_0)^2  \int_{-\infty}^{t_0} dt \int \frac{d^{D-1}\pb}{(2\pi)^{D-1} }  \int [ds_i]\Gamma\left(s_i \pm \frac{\nu}{2}\right) \left(\frac{k}{2}\right)^{-2s_1-2s_2}\left(\frac{p}{2}\right)^{-2s_3-2s_4} \nn\\ \times e^{\pi i (s_1+ s_2 + s_3 - s_4)}(-t)^{D-2-2\sum s_i }\,,  
    \end{eqnarray}
    where $i= 1,2,3,4$. The loop momentum integration is
    \begin{equation}
    \label{eq,loopint}
     \int \frac{d^{D-1}\pb}{(2\pi)^{D-1} } \frac{1}{p^{2s_3 + 2s_4}} = \frac{2\pi^{(D-1)/2} }{(2\pi )^{D-1}\Gamma\left(\frac{D-1}{2}\right)} \int_0^\infty \frac{d p }{p^{2s_3 + 2s_4 -D +2}} = \frac{2\pi^{(D-1)/2} }{(2\pi )^{D-1}\Gamma\left(\frac{D-1}{2}\right)}  2\pi i \delta \left(2s_3 + 2s_4 -D+1 \right) \, ,
    \end{equation}
where, to ensure that the integral is convergent, we require  $Re (-2s_3 -2s_4+D-1 ) <0 $. 
    
    In this diagram, the integration over time and the loop integration are completely separated, and each one contributes a $\delta$-function. The time integration is
    \begin{equation}
        \int_{-\infty}^{t_0} (-t)^{D-2-\sum s_i }   \xrightarrow{\lim_{t_0 \to 0}}  2\pi i \delta \left(D-1 -2\sum s_i\right)\,.
    \end{equation}
    If we were to directly substitute this into the correlation function, and integrate $s_1$ and $s_3$ using the two delta functions, we would find that the left poles and right poles overlap in the $s_2$ integration. We therefore need to introduce a regulating parameter for the time integration as follows,
    \begin{equation}
       (-t_0)^{-2\epsilon} \int_{-\infty}^{t_0} (-t)^{D-2-\sum s_i +2\epsilon}   \xrightarrow{\lim_{t_0 \to 0}}  2\pi i \delta \left(D-1+2\epsilon  -2\sum s_i\right) \,,
    \end{equation}
    where $(-t_0)^{-2\epsilon}$ is introduced to keep the dimension correct. 
The resulting quartic loop contribution to the $2$-point correlation function is then
    \begin{eqnarray}
    && -\frac{i}{2}e^{\frac{\pi i (2D+1 )}{2}} f^\ast_k(t_0)^2 (-t_0)^{-2\epsilon}  \int [ds_i]\Gamma\left(s_i \pm \frac{\nu}{2}\right) \left(\frac{k}{2}\right)^{-2s_1-2s_2} e^{\pi i (s_1+ s_2 + s_3 - s_4)}  \nonumber\\
    &\times & \frac{2\pi^{(D-1)/2} }{(2\pi )^{D-1}\Gamma\left(\frac{D-1}{2}\right)}   \frac{2\pi i \delta \left(2s_3 + 2s_4 -D+1 \right) }{2^{-2s_3-2s_4}}  2\pi i \delta \left(D-1+2\epsilon  -2\sum s_i\right)  \,.
    \end{eqnarray}
Since we have two delta functions here, we can choose to carry out the $s_1$ and $s_3$ integrations. We also deform the space-time dimension $D = 4+ 2\epsilon$ to avoid the overlap of poles in the $s_4$ integration, and obtain
    \begin{eqnarray}
      B_{2q}^{1+}&=&  \frac{e^{2 \pi i \epsilon}}{8}f^\ast_k(t_0)^2 \left(-\frac{k t_0}{2}\right)^{-2\epsilon }   \frac{2\pi^{(D-1)/2} }{(2\pi )^{D-1}\Gamma\left(\frac{D-1}{2}\right)}     \nonumber\\
    &\times &    \int [ds_i] \Gamma\left(s_{2,4} \pm \frac{\nu}{2}\right)  \Gamma\left(\epsilon-s_2 \pm \frac{\nu}{2}\right) \Gamma\left(\frac{3}{2}+\epsilon -s_4 \pm \frac{\nu}{2}\right)    e^{\pi i (\epsilon + \frac{3}{2} - 2s_4)} \,.
    \end{eqnarray}
    Notice that, using Barnes' lemma, the $s_2$ integration yields a $1/\epsilon$-like divergence,
    \begin{equation}
        \int \frac{ds_2}{2\pi i} \Gamma\left(s_2 + \frac{\nu}{2}\right)\Gamma\left(s_2 - \frac{\nu}{2} \right)\Gamma\left(\epsilon-s_2 + \frac{\nu}{2}\right)\Gamma\left(\epsilon-s_2 -  \frac{\nu}{2} \right)  = \frac{\Gamma\left(\epsilon+ \nu\right)\Gamma\left(\epsilon-  \nu\right)\Gamma\left(\epsilon\right)^2   }{\Gamma\left(2\epsilon\right) }\ .
    \end{equation}
   The $s_4$ integration has an extra phase factor, which can not be computed using Barnes' lemma directly. Instead, we close the contour to the left, and enclose two series of poles at $s_2 = -n \pm \frac{\nu}{2}$. Summing over the residues at these two series of poles, we obtain two gaussian hypergeometric functions,
   \begin{eqnarray}
   \label{eq,s4contourint}
         && \int \frac{ds_4}{2\pi i} \Gamma\left(s_4 + \frac{\nu}{2}\right)\Gamma\left(s_4 - \frac{\nu}{2} \right)\Gamma\left(\frac{3}{2}+ \epsilon-s_4 + \frac{\nu}{2}\right)\Gamma\left(\frac{3}{2}+ \epsilon -s_4 -  \frac{\nu}{2} \right) e^{\pi i  \left(\frac{3}{2}+ \epsilon-2 s_4\right)} \nonumber\\
         &=& - e^{\pi i \left(\frac{3}{2} + \epsilon\right)}  \Gamma\left(\frac{3}{2}+ \epsilon \right) \left( \frac{\Gamma\left(\nu\right)\Gamma\left(1- \nu\right) }{\Gamma\left(1+\nu\right)}\Gamma\left(\frac{3}{2}+\nu+  \epsilon \right) ~_2F_1\left(\frac{3}{2}+\epsilon, \frac{3}{2}+\epsilon+\nu; 1+\nu; 1\right) e^{\pi i \nu}  + (\nu \to -\nu)\right) \,.\nn\\
   \end{eqnarray}
   Now, by Gauss' theorem, when $\mathfrak{Re}(c-a-b) >0$, the hypergeometric function becomes
   \begin{equation}
       ~_2F_1\left(a,b,c;1\right) = \frac{\Gamma\left(c\right)\Gamma\left(c-a-b \right) }{\Gamma\left(c-a\right)\Gamma\left(c-b\right)}\ .
   \end{equation}
  We use this to analytically evaluate the integration in the region where $\epsilon $ goes to zero, and the integration \eqref{eq,s4contourint} then becomes
  \begin{eqnarray}
  i\pi \frac{\Gamma\left(-2-2\epsilon\right)}{\Gamma\left(-\frac{1}{2}-\epsilon\right) } \left(\frac{\Gamma\left(-1+ \epsilon\right)}{\Gamma\left(-3 -\epsilon\right)  } + \frac{\Gamma\left(4+\epsilon\right)}{\Gamma\left(2-2\epsilon\right)}\right) =0 \,,
  \end{eqnarray}
    and the quartic loop does not contribute to the 2-point correlator.

\subsection{Two-point reducible diagram}
We next compute the reducible diagram, part c) in figure \ref{figure1}. This diagram contains the same massless loop as in the quartic loop case. Without computation, we know that this loop integration gives a delta function as in \eqref{eq,loopint}, which separates out from the other parts of the computation, and will end up as in \eqref{eq,s4contourint}. We therefore conclude that this massless tadpole loop diagram does not contribute to the UV divergence when computed in dimensional regularization, in the same way as it does not contribute to flat space amplitudes. This also explains why the there is no quantum correction to the one-point function tadpole.
    
       \subsection{Cubic loop}
     Finally we turn to the contribution from part a) of figure \ref{figure1}, the irreducible diagram with 2 cubic vertices. After integrations by parts and using the external lowest order equations of motion, the $1/C^2$ cubic vertex \eqref{cubicvertle3e} reduces to the following form, 
    \begin{equation}
        S_{3}\sim \int d t \frac{ d^{D-1} 
    \kb }{(2 \pi)^{D-1}} (-t)^{\frac{D}{2} -3 } \phi^{3} \,.
    \end{equation}

The terms contributing to the correlator coming from the expansion of \eqref{uintexpe}, which we call $B_{2c}^{(1)++},B_{2c}^{(1)--}, B_{2c}^{(1)-+} $, in which the subscript $c$ indicates that the correction is from cubic loop term, are defined as follows (Notice an extra factor of $1/2$ from the Taylor expansions of $B_{2c}^{(1)++}$ and $B_{2c}^{(1)--}$):
    \begin{eqnarray}
    \label{eq,b2cpp}
        B_{2c}^{(1)++} &\equiv & \left\langle 0 \left|\hat{\phi}\left(k , t_{0}\right) \hat{\phi}\left(k , t_{0}\right) \left(-\frac{1}{2} \int_{-\infty}^{t_{0}} d t_{1} \int_{-\infty}^{t_{0}} d t_{2} \mathcal{H}_{\text {int }}\left(t_{1}\right) \mathcal{H}_{\text {int }}\left(t_{2}\right)  \right) \right| 0\right\rangle\nonumber\\
        &=& -\frac{1}{2} f_k^\ast(t_0)^2 \int_{-\infty}^{t_{0}} d t_{1,2} (-t_1)^{\frac{D}{2}-3}(-t_2)^{\frac{D}{2}-3}  f_k(t_1) f_k(t_2)   \int_{\pb_{1,2}} G_{++}(p_1, t_1, t_2 )G_{++}(p_2, t_1, t_2 )\,,\nonumber
    \end{eqnarray}
    where $\int_{\pb_{1,2}}$ is short for $\int \frac{d^{D-1} \pb_{1,2}}{(2\pi)^{2D-2}}\delta(\pb_1 + \pb_2 + \kb)$. 
The corresponding term $B_{2c}^{--}$ is then given by the complex conjugation of $B_{2c}^{(1)++}$. We also have
     \begin{eqnarray}
    \label{eq,b2cmp}
        B_{2c}^{(1)-+} &\equiv & \left\langle 0 \left| \left((+i )\int_{-\infty}^{t_{0}} d t_{2}   \mathcal{H}_{\text {int }}\left(t_{2}\right)\right)   \hat{\phi}\left(k , t_{0}\right) \hat{\phi}\left(k , t_{0}\right)\left(  (-i)  \int_{-\infty}^{t_{0}} d t_{1}   \mathcal{H}_{\text {int }}\left(t_{1}\right)\right)  \right| 0\right\rangle\nonumber\\
        &=&  f_k(t_0) f_k^\ast(t_0) \int_{-\infty}^{t_{0}} d t_{1,2} (-t_1)^{\frac{D}{2}-3}(-t_2)^{\frac{D}{2}-3}  f_k(t_1) f_k^\ast (t_2)   \int_{\pb_{1,2}} G_{+-}(p_1, t_1, t_2 )G_{+-}(p_2, t_1, t_2 )\,.\nonumber
    \end{eqnarray} 
    
    We start with $B_{2c}^{(1)++}$. Expanding the bulk-to-bulk propagator, we have
     \begin{equation}
         G_{++}\left(p_{1}, t_{1}, t_{2}\right) G_{++}\left(p_{2}, t_{1}, t_{2}\right) = \theta(t_1 - t_2 ) f_{p_1}^\ast (t_1 )f_{p_2}^\ast (t_1 ) f_{p_1} (t_2 ) f_{p_2} (t_2 )  + (t_1 \leftrightarrow t_2 )\,.
     \end{equation}
     For the two different time orderings, we denote the one containing $\theta(t_1 - t_2)$ by $B_{2c}^{(1)\pm\pm, >}$ and the one containing $\theta(t_2 - t_1)$ by $B_{2c}^{(1)\pm\pm, <}$.  

     Using the Mellin-Barnes representation for the mode functions \eqref{MBrepmodefunctionsee}, we have
    \begin{eqnarray}
     && B_{2c}^{(1)++,>} = -  \frac{e^{\frac{\pi i (2D+1 )}{2}}  f_k^\ast( t_0)^2 }{(4\pi)^3} \int_{-\infty}^{t_0} dt_1 \int_{-\infty}^{t_1} dt_2 \int [ds_i] \Gamma\left(s_i + \frac{\nu_i}{2}\right)\Gamma\left(s_i - \frac{\nu_i}{2}\right) e^{\pi i (s_1 + s_2 +s_5 +s_6 - s_3 -s_4)} \nonumber\\
      &  & \left(\frac{ k}{2}\right)^{-2(s_1 + s_2)} \int \frac{d^{D-1} \pb_{1 }}{(2\pi)^{D-1}} (-t_1)^{-2(s_1+s_3+s_4)+\frac{D-3}{2} } (-t_2)^{-2(s_2+s_5+s_6)+\frac{D- 3}{2}} \left(\frac{ p_1}{2}\right)^{-2(s_3 + s_5)}\left(\frac{ |p_1+ k|}{2}\right)^{-2(s_4 + s_6)} \ , \nn\\ 
      \label{eq:B++>}
    \end{eqnarray}
    and
     \begin{eqnarray}
     && B_{2c}^{(1)++,<} = -    \frac{e^{\frac{\pi i (2D+1 )}{2}} f_k^\ast( t_0)^2 }{(4\pi)^3} \int_{-\infty}^{t_0} dt_1 \int_{t_1}^{t_0} dt_2 \int [ds_i] \Gamma\left(s_i + \frac{\nu_i}{2}\right)\Gamma\left(s_i - \frac{\nu_i}{2}\right) e^{\pi i (s_1 + s_2+s_3 +s_4 -s_5 -s_6 )} \nonumber\\
      &  & \left(\frac{ k}{2}\right)^{-2(s_1 + s_2)} \int \frac{d^{D-1} \pb_{1 }}{(2\pi)^{D-1}} (-t_1)^{-2(s_1+s_3+s_4)+\frac{D-3}{2} } (-t_2)^{-2(s_2+s_5+s_6)+\frac{D- 3}{2}} \left(\frac{ p_1}{2}\right)^{-2(s_3 + s_5)}\left(\frac{ |p_1+ k|}{2}\right)^{-2(s_4 + s_6)}
            \label{eq:B++<} \, ,\nn\\
    \end{eqnarray} 
    where $\int [ds_i] \equiv \Pi_{i=1}^6 \int_{c- i\infty}^{c+i \infty} ds_i /2\pi i$, and we have used momentum conservation to write $\pb_2 = \pb_1 + \kb$. 
    
 Note that both of the above two expressions have the same momentum integrals. We now use the Schwinger parameterization:
    \begin{eqnarray} 
        && \int \frac{d^{D-1} \pb}{(2\pi)^{D-1} } \frac{1}{|p|^{2s_3+2s_5}|p+ k|^{2s_4+2s_6}} \nonumber\\
        &=&   \frac{k^{ D-1-2(s_3-s_4 -s_5 -s_6)  } \Gamma\left(s_3+s_4+s_5+s_6 - \frac{D-1}{2} \right) \Gamma\left(\frac{D-1}{2}-s_3 -s_5 \right)\Gamma\left(\frac{D-1}{2}-s_4 -s_6 \right) }{(4\pi)^{\frac{D-1}{2} }\Gamma\left( s_3 +s_5 \right)\Gamma\left( s_4 +s_6 \right)   \Gamma\left(D-1- (s_3+s_4+s_5+s_6 )   \right) } \ \nn\\
    \end{eqnarray}
    to perform these integrals. This standard result introduces a degeneracy in the poles of the integrand; when evaluating the residue at the pole at $s_3+s_5 = \frac{D-1}{2} $, due to the term $\Gamma\left(\frac{D-1}{2} - s_3 -s_5\right)$, we find a cancellation between the term $\Gamma\left(s_3 + \cdots s_6 - \frac{D-1}{2} \right)$ in the numerator, and the term $\Gamma(s_4 +s_6)$ in the denominator. Therefore, when shifting the contour, it is possible that we might miss some of the necessary poles because the order in which we choose to shift the contour depends on the arbitrary ordering of the variables. To take account of this possibility, following \cite{Premkumar:2021mlz}, we introduce a new parameter that allows us to ensure that the pole structure is manifest at all times. 
   
   Treating the momentum integral as a convolution of two copies of $\tilde f_s(p) \equiv 1/p^s $, for different values of $s$, we have
   \begin{eqnarray}
    &&  \int \frac{d^{D-1} \mathbf{p}}{(2 \pi)^{D-1}} \frac{1}{|p|^{2 s_{3}+2 s_{5}}|p+k|^{2 s_{4}+2 s_{6}}} =  \int \frac{d^{D-1} \mathbf{p}}{(2 \pi)^{D-1}}  \tilde f_{2s_3 + 2s_5}(\pb) \tilde f_{2s_4 + 2s_6}(\pb+\kb) \nonumber\\
     && =   \int  d^{D-1} \mathbf{y}\  e^{- i \kb \cdot \mathbf{y} }  f_{2s_3 + 2s_5} (\mathbf{y})  f_{2s_4 + 2s_6 } (\mathbf{y})  \ ,
   \end{eqnarray}
   where $f_s(\mathbf{y}) = \int \frac{d^{D-1} \mathbf{p}}{(2 \pi)^{D-1}}  e^{i \pb \cdot \mathbf{y} } \tilde f_s (p) = \frac{1}{2^{s} \pi^{(D-1)/ 2}} \frac{\Gamma\left(\frac{D-s}{2}\right)}{\Gamma\left(\frac{s}{2}\right)} y^{s-D} $. We then regularize the dimension of $\mathbf{y}$ from $D$ to $\bar D$, and include an extra factor of $(- t_0)^{D-\bar D} $ in order to ensure that the overall dimension remains correct. 
     \begin{eqnarray}
    \label{eq,loopdiv}
        && \int \frac{d^{D-1} \pb}{(2\pi)^{D-1} } \frac{1}{|p|^{2s_3+2s_5}|p+ k|^{2s_4+2s_6}} = (-t_0)^{D-\bar D }\int  d^{\bar D-1} \mathbf{y}\  e^{- i \kb \cdot \mathbf{y} }  f_{2s_3 + 2s_5} (\mathbf{y})  f_{2s_4 + 2s_6 } (\mathbf{y}) \nonumber\\
        &=&   \frac{k^{ D-1-2(s_3-s_4 -s_5 -s_6)  } \Gamma\left(s_3+s_4+s_5+s_6 - \frac{2D-\bar{D} -1 }{2} \right) \Gamma\left(\frac{D-1}{2}-s_3 -s_5 \right)\Gamma\left(\frac{D-1}{2}-s_4 -s_6 \right) }{(4\pi)^{\frac{2D-\bar{D} -1}{2} } (-k t_0)^{\bar{D}- D }\Gamma\left( s_3 +s_5 \right)\Gamma\left( s_4 +s_6 \right)   \Gamma\left(D-1- (s_3+s_4+s_5+s_6 )   \right) } \,.\nn\\
    \end{eqnarray}
The parameter $D- \bar D $ now allows us more freedom when choosing the contour, and we are thus able to avoid the ambiguities discussed above. In contrast with the case of  the quartic loop, in which integration over the loop momentum yields a delta function, directly enforcing a conservation law, in the cubic case the loop integration leads to more complicated pole structures. The difference can be understood similarly to what happens in flat space, where the massless quartic loop is a vacuum massless bubble and does not contribute to the self-energy when computed using dimensional regularization. In our de Sitter case, for the specific example above, we can regulate the cubic loop appropriately by choosing $D = 4 + 2\kappa$ and $\bar D = 4$.  
   
 We now turn to the time integrations in our expressions \eqref{eq:B++>} and \eqref{eq:B++>}. Focusing first on the time integration in \eqref{eq:B++>}, we obtain: 
    \begin{equation}
        \int_{-\infty}^{t_0} dt_1\int_{-\infty}^{t_1} dt_2 (-t_1)^{\alpha} (-t_2)^{\beta} = \int_{-\infty}^{t_0} dt_1  \frac{(-t_1)^{\alpha+\beta+ 1}  }{ \beta+ 1 } \ .
    \end{equation}
We then take the late time limit $t_0 \to 0$, so that the integration over $t_1$ becomes a $\delta$ function, giving
    \begin{equation}
        \lim_{t_0 \to 0} \int_{-\infty}^{t_0} dt_1  \frac{(-t_1)^{\alpha + \beta+1 }  }{ \beta+1 }  = \frac{2 i \pi \delta\left( \alpha+\beta +2 \right)}{\beta + 1 }\ ,
    \end{equation}
    where $\alpha = -2s_1 -2s_3 -2s_4 + \frac{D-3}{2}$, $\beta = -2s_2 -2s_5 -2s_6 + \frac{D-3 }{2} $, and we require $\mathfrak{Re}(\beta) < -1$ to ensure that the $t_2$ integration converges. If we were to directly integrate out $s_1$ using the $\delta$ function, we would encounter the same ambiguity that we found when evaluating the quartic loop case. We therefore once again regulate by introducing a parameter $\epsilon$,
    \begin{eqnarray}
    \label{eq,regulat}
    && (-t_0)^{-4 \epsilon   } \int_{-\infty}^{t_0} dt_1\int_{-\infty}^{t_1} dt_2 (-t_1)^{\alpha+ 2\epsilon} (-t_2)^{\beta +2\epsilon} 
    \xrightarrow{\lim_{t_0 \to 0}}   \frac{2 i \pi \delta\left(\alpha+ \beta +4\epsilon + 2 \right)}{\beta+2 \epsilon+1 } \ .
    \end{eqnarray}
 This converges if we require the real part of denominator to be negative, $\mathfrak{Re}(\beta+ 2\epsilon+1) <0$. Since $\beta = -2s_2-2s_5-2s_6$, this requires us to choose the contour to the right of the pole in the denominator. 
        
    In a similar way, if we now focusing instead on the time integration in \eqref{eq:B++<}, we obtain:
    \begin{eqnarray}
    \label{eq,regulat}
    && (-t_0)^{-4 \epsilon   } \int_{-\infty}^{t_0} dt_2\int_{-\infty}^{t_2} dt_1 (-t_1)^{\alpha+ 2\epsilon} (-t_2)^{\beta +2\epsilon} 
    \xrightarrow{\lim_{t_0 \to 0}}   \frac{2 i \pi \delta\left(\alpha+ \beta +4\epsilon + 2 \right)}{\alpha+2 \epsilon+1 } \ ,
    \end{eqnarray} 
    and to ensure that the $t_1$ integration is convergent as $t_1 \to -\infty$, we require that the denominator  $\mathfrak{Re}(\alpha+2\epsilon + 1 ) < 0$.

Using these results, we then have 
    \begin{eqnarray}
    && B_{2c}^{(1)++,>} = - \frac{ e^{\frac{\pi i (2D+1 )}{2}} f_k^\ast(t_0)^2  k^{D-1}  (-k t_0)^{2\kappa}}{(4\pi)^{3+ (D-1)/2+ \kappa }}  \int   [ds_i] \Gamma\left(s_i \pm \frac{\nu_i}{2}\right)  e^{\pi i (s_1 + s_2 +s_5 +s_6 - s_3 -s_4)} \left(\frac{ k}{2}\right)^{-2(s_1 + \cdots +s_6)}
      \nonumber\\
      &\times &   \frac{2 i \pi \delta(\alpha+\beta+2+ 4\epsilon ) (-t_0)^{-4\epsilon}}{\beta+2\epsilon +1 }    \frac{  \Gamma\left(s_3+s_4+s_5+s_6 - \frac{D-1}{2} \right) \Gamma\left(\frac{D-1}{2}-s_3 -s_5 \right)\Gamma\left(\frac{D-1}{2}-s_4 -s_6 \right) }{ \Gamma\left( s_3 +s_5 \right)\Gamma\left( s_4 +s_6 \right)   \Gamma\left(D-1 - (s_3+s_4+s_5+s_6 )   \right) }   \,,\nn\\
    \end{eqnarray}
    where $i = 1,\cdots , 6$. Integrating $s_1$ with respect to the $\delta$ function then yields
    \begin{eqnarray}
    \label{eq,b2cppg}
    && B_{2c}^{(1)++,>} = - \frac{e^{\frac{\pi i (2D+1 )}{2}} f_k^\ast(t_0)^2    }{(4\pi)^{3+ \frac{D-1}{2} } \pi^\kappa  }  \left(- \frac{ k t_0}{2}\right)^{2\kappa -4 \epsilon }   \int   [ds_i] \Gamma\left(s_i \pm \frac{\nu_i}{2}\right)    \Gamma\left(\frac{3 }{2}+\kappa+ 2 \epsilon - \sum_{i=2}^6  s_i \pm  \frac{\nu}{2}\right)
      \nonumber\\
      &\times &  \frac{e^{\pi i \left(\frac{3}{2}+\kappa+ 2 \epsilon - 2s_3 -2 s_4\right)}  }{2 (\beta+2\epsilon+ 1)}   \frac{  \Gamma\left(s_3+s_4+s_5+s_6 - \frac{3+4\kappa }{2} \right) \Gamma\left(\frac{3+2\kappa}{2}-s_3 -s_5 \right)\Gamma\left(\frac{3+2\kappa }{2}-s_4 -s_6 \right) }{  \Gamma\left( s_3 +s_5 \right)\Gamma\left( s_4 +s_6 \right)   \Gamma\left(3+2\kappa  - (s_3+s_4+s_5+s_6 )   \right) }    \ , \nn\\
    \end{eqnarray}

In the same way, we obtain
     \begin{eqnarray}
      \label{eq,b2cppl}
    && B_{2c}^{(1)++,<} = - \frac{e^{\frac{\pi i (2D+1 )}{2}} f_k^\ast(t_0)^2    }{(4\pi)^{3+ \frac{D-1}{2} } \pi^\kappa }  \left(- \frac{ k t_0}{2}\right)^{2\kappa -4 \epsilon }   \int   [ds_i] \Gamma\left(s_i \pm \frac{\nu_i}{2}\right)    \Gamma\left(\frac{3 }{2}+\kappa+ 2 \epsilon - \sum_{i=2}^6  s_i \pm  \frac{\nu}{2}\right)
      \nonumber\\
      &\times &  \frac{e^{\pi i \left(\frac{3}{2}+\kappa+ 2 \epsilon - 2s_5 -2 s_6\right)}  }{ - 2 (\beta+2\epsilon+ 1)}   \frac{  \Gamma\left(s_3+s_4+s_5+s_6 - \frac{3+4\kappa }{2} \right) \Gamma\left(\frac{3+2\kappa}{2}-s_3 -s_5 \right)\Gamma\left(\frac{3+2\kappa }{2}-s_4 -s_6 \right) }{  \Gamma\left( s_3 +s_5 \right)\Gamma\left( s_4 +s_6 \right)   \Gamma\left(3+2\kappa  - (s_3+s_4+s_5+s_6 )   \right) }    \ ,\nn\\
    \end{eqnarray} 
    where $i = 2,\ldots ,6$.  The denominator in $B_{2c}^{(1)++,<} $ arises from integrating over the $\delta$-function $\delta \left(\alpha+\beta+4\epsilon+2\right)$, and the convergence of the $t_1$ integral now requires that $-\mathfrak{Re} \left(\beta + 2\epsilon+1\right) =\mathfrak{Re}\left( \alpha+2\epsilon+1  \right) < 0 $.  We therefore choose the contour to run to the left of the pole. 
    
   Notice that the contour integration itself does not involve any momentum dependence. As a result, the momentum structure of the cubic loop correction has the late time limit $\mathfrak{Re} \left(f_k^\ast (t_0)^2 \right) $, which is of the same form as the higher derivative term that is proportional to $c_4^\prime$. One can also check that in the late time limit, we have $f_k(t_0)^2=f_k^\ast(t_0)^2 =f_k(t_0)f_k^\ast(t_0)  $. We can therefore directly sum the divergent terms, since each of $B_{2c}^{(1)++}, B_{2c}^{(1)--} $ and $B_{2c}^{(1)-+}$ have the same momentum dependence. What remains is to compute the divergences in the contour integration that is independent of the momentum. 
   
 As discussed in \cite{Premkumar:2021mlz}, the divergences arise when the left poles overlap with the right poles. To evaluate these divergences, we introduce regulating parameters $\epsilon_i $ that allow us to choose fixed contour lines that are parallel to the imaginary axis and that separate the left poles from the right poles, and then analytically continue the result to the region $\epsilon_i \approx 0 $. This procedure has been carefully discussed in \cite{Tausk:1999vh, Czakon:2005rk, Premkumar:2021mlz}. We use the above regularization and choose $\kappa = \epsilon$, and employ the Mathematica package MB.m \cite{Czakon:2005rk} to yield the following divergence structure, where we have summed over $B_{2c}^{(1)++} =B_{2c}^{(1)++,>} + B_{2c}^{(1)++,<} $ and $B_{2c}^{--} = B_{2c}^{(1)--,>} + B_{2c}^{(1)--,<}$,
    \begin{equation}
        \label{eq,cubicdiv}
      \left( -\frac{1}{360 \sqrt{\pi} \epsilon^2}  +  \frac{\alpha_1  }{\epsilon}    \right) \left(- \frac{kt_0}{2}\right)^{-2\epsilon} f_k(t_0) \,f_k^\ast(t_0)\,,
    \end{equation}
 with $\alpha_1$ a constant. We provide more details of the contour choices in the Appendix.

    \subsubsection{$B_{2c}^{(1) -+}$}
    For the $B_2^{(1) -+}$ correlator, both time integrations start from $-\infty$ and end at $t_0$, and the loop divergence is same as in Eq.\eqref{eq,loopdiv}:
    \begin{eqnarray}
      B_{2c}^{(1) -+} &=& f_k(t_0) f_k^\ast(t_0)  \int _{-\infty}^{t_0} dt_{1,2}    \Pi_{i =1 }^ 6 \int [ds_i] \Gamma\left(s_i \pm \frac{\nu_i}{2}\right) e^{\pi i (s_1 + s_3 +s_4 -s_2 - s_5 -s_6)}  \left(\frac{ k}{2}\right)^{-2(s_1 + s_2)} \nonumber\\
      &\times & \int \frac{d^{D-1} \pb_{1 }}{(2\pi)^{D-1}}(-t_1)^{-2(s_1+s_3+s_4)+\frac{D-3}{2}} (-t_2)^{-2(s_2+s_5+s_6)+\frac{D-3}{2}} \left(\frac{ p_1}{2}\right)^{-2(s_3 + s_5)}\left(\frac{ |p_1+ k|}{2}\right)^{-2(s_4 + s_6)} \ . \nn\\
    \end{eqnarray}
 Performing the time integrations, and taking the late time limit $t_0 \to 0$, we introduce the same regulating parameters as in \eqref{eq,regulat} to obtain
    \begin{equation}
     (-t_0)^{-4 \epsilon   } \int_{-\infty}^{t_0} dt_1\int_{-\infty}^{t_0} dt_2 (-t_1)^{\alpha+ 2\epsilon} (-t_2)^{\beta +2 \epsilon} 
    \xrightarrow{\lim_{t_0 \to 0}}   2 i \pi \delta\left(\alpha  +2\epsilon + 1 \right)  2 i \pi \delta\left(  \beta +2 \epsilon + 1 \right) \ ,
     \end{equation}
     where again (as in the $++$ case), $\alpha = -2s_1 -2s_3 -2s_4 + \frac{D-3}{2}$, $\beta = -2s_2 -2s_5 -2s_6 + \frac{D-3 }{2} $, and we require the real parts of $\alpha+2\epsilon+ 1 $ and $\beta +2\epsilon +1$ to be negative to converge.

     Using the $\delta$-functions to perform the integrals over $s_1$ and $s_2$, the $+-$ correlator is then
     \bea
      && B_{2c}^{(1) -+} = \frac{f_k(t_0) f_k^\ast(t_0)}{4 (4\pi)^{\frac{D-1}{2} } \pi^\kappa   }    \left(-\frac{ k t_0 }{2}\right)^{2\kappa -4\epsilon }   \Pi_{i =3 }^ 6 \int [ds_i] \Gamma\left(s_i \pm \frac{\nu_i}{2}\right)\Gamma\left(\frac{3+2\kappa} {4}+ \epsilon - s_3 -s_4  \pm \frac{\nu}{2}\right) \nonumber\\
      && \Gamma\left(\frac{ 3+2\kappa} {4}+ \epsilon  - s_5 -s_6  \pm \frac{\nu}{2}\right)    \frac{  \Gamma\left(s_3+s_4+s_5+s_6 - \frac{3+4\kappa}{2} \right) \Gamma\left(\frac{3+2\kappa}{2}-s_3 -s_5 \right)\Gamma\left(\frac{3+2\kappa}{2}-s_4 -s_6 \right) }{  \Gamma\left( s_3 +s_5 \right)\Gamma\left( s_4 +s_6 \right)   \Gamma\left(3+2\kappa - (s_3+s_4+s_5+s_6 )   \right) }     \ . \nn\\
     \eea
     Finally, choosing $\kappa = \epsilon$ and applying the technique described earlier, we can see that the divergent part of this expression takes the form
      \begin{equation} 
      \left( \frac{1}{360 \sqrt{\pi} \epsilon^2}  +  \frac{\alpha_2  }{\epsilon}    \right) \left(- \frac{kt_0}{2}\right)^{-2\epsilon} f_k(t_0) f_k^\ast(t_0)\ . 
    \end{equation}

With the divergent parts of the $B_{2c}^{(1) ++}$, $B_{2c}^{(1) --}$, and $B_{2c}^{(1) -+}$ correlators in hand, we can now sum them and observe that the $\frac{1}{\epsilon^2} \left(-\frac{k t_0}{2}\right)^{-2\epsilon}$ divergences cancel, leaving only $\frac{1}{\epsilon} \left(-\frac{k t_0}{2}\right)^{-2\epsilon}$-type divergences.  As in flat spacetime, we can absorb the  $1/\epsilon$ divergence into the coefficient $c_{4}^\prime$ and treat the higher derivative term as a counter term. Thus, the 1-loop correction to the 2-point equal time correlation function in the late time limit is
    \begin{equation}
        \left\langle\phi(t_0, \mathbf{k}) \phi\left(t_0, \mathbf{k}^{\prime}\right)\right\rangle_{\text{1-loop}}= f_{\mathbf{k}}(t_0) f_{\mathbf{k}}^{*}\left(t_0 \right) \log\left(- \frac{k t_0}{2}\right)\,.
    \end{equation}
    
Using this, the full equal time correlator $\braket{\Phi(t_0, \mathbf{x}), \Phi(t_0, \mathbf{x}^\prime)}$ is therefore given by,
 \begin{eqnarray}
&&    \left\langle\Phi(t_0, \mathbf{k}) \Phi\left(t_0, \mathbf{k}^{\prime}\right)\right\rangle \nn\\
    &&= \frac{C^{2 \Delta}}{(-t_0)^{2\Delta} }\left[1+\frac{1}{C^{D-2}}(-t_0)^{D-2} \left(\left\langle\phi(t_0, \mathbf{k}) \phi\left(t_0, \mathbf{k}^{\prime}\right) \right\rangle_{\text{0-loop}} + \left\langle\phi(t_0, \mathbf{k}) \phi\left(t_0, \mathbf{k}^{\prime}\right) \right\rangle_{\text{1-loop}} + \cdots\right) \right]\nonumber\\
    &&=  \frac{C^{2 \Delta}}{(-t_0)^{2\Delta} }\left[1+\frac{f_k(t_0) f_k^\ast(t_0)}{C^{D-2}}(-t_0)^{D-2} \left(1+ \frac{const.}{C^2} \log\left(-\frac{k t_0}{2}\right) + \cdots\right) \right] \,.
 \end{eqnarray}
Now, taking the late-time limit of the mode functions yields
  \begin{equation}
     f_k(t_0) f_k^\ast(t_0) = \frac{(-t_0) \left(-\frac{1}{2} k t_0\right)^{-D-1}    }{4\pi \sin\left(\frac{D+1}{2}\pi\right)^2 \Gamma\left(\frac{1-D }{2}\right)^2 } \sum_{m=0} c_m \left( -   k^2 t_0^2 \right)^m\ ,   c_m =  \frac{\left(-\frac{D}{2} \right)_m}{m! \left(\frac{1- D}{2} \right)_m \left( - D  \right)_m   }  \ ,
  \end{equation}
   where $(a)_m = a(a+1)(a+2)\cdots (a+m -1)$ is the Pochhammer symbol.  Furthermore, using the Fourier identity  
  \begin{eqnarray}
     \int \frac{ d^{d} \mathbf{k} }{(2\pi)^d} \frac{e^{i \mathbf{k} \cdot  \left(\mathbf{x}-\mathbf{x}^\prime \right) }}{\mathbf{k}^{2 \alpha}}= B_{d}(2 \alpha) \left(\frac{|\mathbf{x}-\mathbf{x}^\prime |}{2} \right)^{2 \alpha-d}, \quad B_{d}(2 \alpha)=\frac{\pi^{d / 2}  }{(2\pi)^d  } \frac{\Gamma\left(\frac{d-2 \alpha }{2}\right)}{\Gamma(\alpha )} \ ,
  \end{eqnarray}
  we obtain 
  \begin{eqnarray}
  \left\langle\phi\left(t_{0}, \mathbf{x}\right) \phi\left(t_{0}, \mathbf{x}^{\prime}\right)\right\rangle_{0-\mathrm{loop}} = \frac{(-t_0) |\mathbf{x}-\mathbf{x}^\prime|^{2} }{4\pi \sin\left(\frac{D+1}{2}\pi\right)^2\Gamma\left(\frac{1-D }{2}\right)^2  } \left(-\frac{1}{2}  t_0\right)^{-D-1} \sum_{m=0} c_m B_d(2\alpha)    \left( \frac{ 4 t_0^2 }{|\mathbf{x}-\mathbf{x}^\prime|^2}   \right)^m  \ , \nn\\
  \end{eqnarray}
  with $2\alpha = D+1 -2m$ and $d = D-1$. Fom this, one can see that in coordinate space the correlator is a function of the cross ratio $\xi$ as defined in \eqref{eq,crossratio}. 
 
Therefore, we finally obtain 
\begin{equation}
    \left\langle\Phi\left(t_{0}, \mathbf{x}\right) \Phi\left(t_{0}, \mathbf{x}^{\prime}\right)\right\rangle_{0 -\text{loop}} = \frac{C^{2\Delta}}{(-t_0)^{2\Delta}} \left[1+ \frac{2^{D+3}}{C^{D-2}}\xi  \sum_{m=0} c_m B_d(2\alpha ) \left(\frac{1}{\xi}\right)^{m}\right]\ ,
\end{equation}
which is consistent with the form of the correlator for a defect CFT as in \eqref{eq,crossratio}. The effective field theory approach thus allows us to read off the arbitrary function $F(\xi)$ to any given order. 

For the 1-loop correlator, the logarithmic contribution can be viewed as the next-to-leading order expansion of $  \left(-\frac{k t_0}{2}\right)^{\epsilon} $. Thus, the loop contribution after performing a Fourier transform would take a similar form as the tree level expansion, with a $1/C^2$ suppression and some changes to the coefficients.

    \section{Conclusions}

    We have explored some systematics of the effective field theory of conformal symmetry breaking, which describes fluctuations around symmetry-breaking VEVs. With this effective theory, correlators can be computed systematically as an expansion in inverse powers of the symmetry-breaking parameter. 
    
   As an example, we began with the computation of the leading terms in the EFT expansion of the 2 point function in a theory in which  conformal symmetry in broken to Poincar\'e symmetry.  The EFT gives an expansion which is valid at long distances, complementary to the short distance operator product expansion.  
      
   We then studied the breaking that occurs due to a space-like defect in a Lorentzian CFT; a setup that is relevant for early universe scenarios in which the reheating surface can be thought of as a conformal defect. We expanded the EFT around the time-dependent symmetry-breaking VEV and computed the correlation functions of the fluctuations. 
   Because the background is time-dependent, we computed the in-in correlators, which involve time integrations from the infinite past to the space-like defect surface. To compute the next-to-leading order corrections to the correlators, one must evaluate the loop diagrams from cubic and quartic interactions. The time integrations in these loop diagrams are in principle extremely complicated, and so we found it convenient to make use of Mellin-Barnes representations for the mode functions. On the CFT side, this symmetry-breaking pattern demands that the general form of the 2-point correlator takes the form of an arbitrary function $F$ of the cross-ratio $\xi$. By  Fourier transforming, we were able to use our EFT expansion to provide a series of terms for this arbitrary function . 
   
While we have demonstrated the usefulness of this technique in specific examples, the idea should be generalizable in a number of ways. For example, although we limited our calculations to 2-point functions, the EFT can in principle by used to compute any correlator in a long distance expansion.  In particular, it would be interesting to compute the 3-point correlator in this our example, which would provide information about non-gaussianity in the pseudo-conformal universe model. Similarly, while we chose to apply our methods to the example of a spacelike defect in a Lorentzian CFT,  other maximally symmetric defects in various signatures should work similarly. It would be interesting to extend our calculations to the case of higher co-dimension defects, and perhaps even to the case in which there is mixing between space-like and time-like defects. Another interesting extension would be to the case with multiple fields $\Phi_i$ acquiring VEVs $\Phi_i=C_i^\Delta/(-t)^\Delta$, where spectator fields would live in different backgrounds and yield different power spectrums. Finally, it may also be worth exploring the subtleties in the special cases $\Delta=0$ and $D\leq 2$ that we have bypassed in this work.

    \vspace{-.2cm}
    \paragraph{Acknowledgments:} We would like to thank Austin Joyce for helpful conversations and comments. QL would like to thank Donggang Wang for helpful suggestions at early stage, and Sam S.C. Wong for useful discussion. KH acknowledges support from DOE grant DE-SC0009946 and from Simons Foundation Award Number 658908. The work of QL and MT is supported in part by US Department of Energy (HEP) Award DE-SC0013528. MT was also supported in part by the Simons Foundation Origins of the Universe Initiative, grant number 658904.
    
    \appendix

    \section{Details of the contour choice} 
    Here we provide more details of how to obtain Eq. \eqref{eq,cubicdiv}.  For a given contour integral over $N$ variables $s_i$,
   \begin{eqnarray}
  \int \frac{ds_1}{2\pi i} \cdots  \int \frac{ds_N}{2\pi i} \prod_i \frac{\Gamma\left(a_i + \sum_k b_{ik} s_k   \right)}{\Gamma\left(c_i + \sum_k d_{ik} s_k \right)}\ ,
   \end{eqnarray}
   the signs of $b_{ik}$ determine whether the Gamma functions result in left poles or right poles. We choose the contours for the above integrals as fixed straight lines that are parallel to the imaginary axis and that intersect with the real axis at $\mathcal{C}_i$. These well-defined contours should separate the left poles and the right poles, which is equivalent to requiring
   \begin{eqnarray}
   \label{eq,condition}
   a_i + \sum_k b_{ik } \mathcal{C}_k > 0\ , \forall i  \ .
   \end{eqnarray}
   
   However, this may not be the case, and we may be faced with a situation in which the left poles overlap with the right poles, and we can no longer define the contour in this simple way. In this case, we can introduce a set of regulating parameters $\epsilon_l$ into the Gamma functions, 
   \begin{eqnarray}
  \int \frac{ds_1}{2\pi i} \cdots  \int \frac{ds_N}{2\pi i} \prod_i \frac{\Gamma\left(a_i + \sum_k b_{ik} s_k   + \sum_l e_{il}\epsilon_l \right)}{\Gamma\left(c_i + \sum_k d_{ik} s_k +\sum_l f_{il }\epsilon_l \right)}\ ,
   \end{eqnarray}
 so that the conditions \eqref{eq,condition} are now satisfied, and then perform an analytic continuation back to $\epsilon_l \rightarrow 0$ afterwards.

\begin{figure}[h!]
\centering
\includegraphics[scale=0.2]{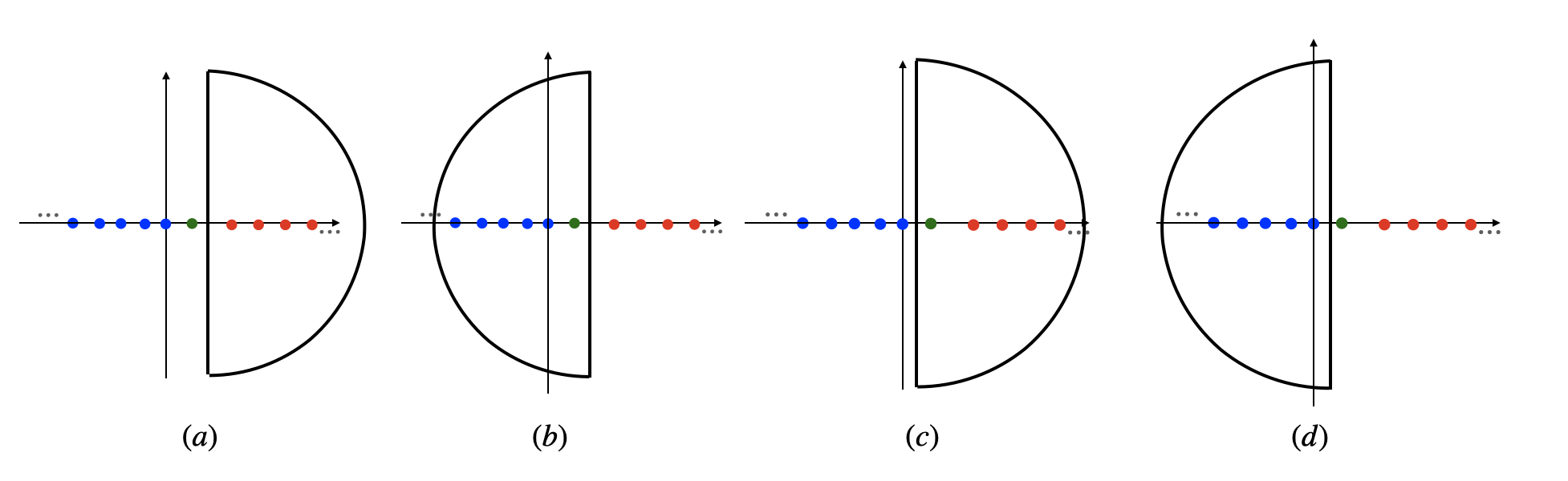} 
\caption{The different contour choices. The green dot represents the simple pole in the denominator, the red dots represent the right poles arising from $\Gamma(a_i + \sum_k b_{ik}s_k)$ when $b_{ik}$ is negative, and the blue dots represent the left poles when $b_{ik}$ is positive. All the poles lie on the real axis. $B_{2c}^{(1)++,>}$ requires the contour line to lie to the right of the simple pole, as shown in $(a)$ and $(b)$. The $B_{2c}^{(1)++,<}$ integral requires the contour line to lie to the left of the simple pole, as shown in $(c)$ and $(d)$.  } 
\label{fig:pole}
\end{figure}

    This analytic continuation procedure is implemented in the Mathematica package MB.m available from \cite{Czakon:2005rk}. Unfortunately, we cannot apply the Mathematica package directly to our contour integrals, e.g. \eqref{eq,b2cppg}, because in our case we have an extra single pole in the denominator arising from the time integration. The convergence of the time integration requires that $s_2 + s_5 + s_6 - \frac{3}{4} -\frac{3 \epsilon}{2} > 0$ in \eqref{eq,b2cppg}, and $s_2 + s_5 + s_6 - \frac{3}{4} -\frac{3 \epsilon}{2} < 0$ in \eqref{eq,b2cppl}. This changes the locations of the contour lines $\mathcal{C}_i$. Since there is no momentum dependence in the contour integral, we are free to close the contour to either the right side or the left side. However, as shown in figure \ref{fig:pole} $(a)$ and $(b)$, taking \eqref{eq,b2cppg} as an example, closing the contour to the left does not encircle the single denominator pole, while closing the contour to the right does include it. In principle, one can choose to close the contour on either side, but the convention of the Mathematica package is to always choose to close the contour to the right. Therefore, the expression for $B_{2c}^{(1)++,>}$ should not involve the residue at this single pole, while the expression for $B_{2c}^{(1)++,<}$ should contain it. In other words, the actual contour choices are those shown in Fig.\eqref{fig:pole} $(a)$ and $(c)$. 
    
    Moreover, since the algorithm is not written to recognize this single pole, it is necessary to first rewrite it as
    \begin{equation}
    \label{eq,fictitious}
        \frac{1}{-2s_2-2s_5-2s_6 + \frac{3}{2} +  3\epsilon  } = \frac{\Gamma\left( -2 \left(\frac{3}{4} + \frac{3\epsilon}{2} -s_2-s_5-s_6 \right) \right)}{\Gamma\left( -2 \left(\frac{3}{4} + \frac{3\epsilon}{2} -s_2-s_5-s_6 \right)+ 1 \right)}\ ,
    \end{equation}
    which generates extra fictitious poles at  $-2\left(\frac{3}{4} + \frac{3\epsilon}{2} -s_2-s_5-s_6 \right) = - n$ for non-negative integers $n$ from the numerator in \eqref{eq,fictitious}. One must then remove these fictitious terms by subtracting the residue at $1/(-2s_2-2s_5-2s_6+ \frac{3}{2} +  3\epsilon )$,
    \begin{eqnarray}
    \label{eq,b2cppodot}
       &&  B_{2c}^{(1)++, >\odot} = -\frac{e^{\frac{\pi i(2 D+1)}{2}} f_{k}^{*}\left(t_{0}\right)^{2}}{(4 \pi)^{3+\frac{D-1}{2}} \pi^{\kappa}}\left(-\frac{k t_{0}}{2}\right)^{2 \kappa-4 \epsilon} \int\left[d s_{i}\right] \Gamma\left(s_{i} \pm \frac{\nu_{i}}{2}\right) \Gamma\left(\frac{3+2 \kappa}{4}+\epsilon-s_{3}-s_{4} \pm \frac{\nu}{2}\right)\nonumber\\
        &\times &\Gamma\left(\frac{3+2 \kappa}{4}+\epsilon-s_{5}-s_{6} \pm \frac{\nu}{2}\right) \frac{\Gamma\left(s_{3}+s_{4}+s_{5}+s_{6}-\frac{3+4 \kappa}{2}\right) \Gamma\left(\frac{3+2 \kappa}{2}-s_{3}-s_{5}\right) \Gamma\left(\frac{3+2 \kappa}{2}-s_{4}-s_{6}\right)}{\Gamma\left(s_{3}+s_{5}\right) \Gamma\left(s_{4}+s_{6}\right) \Gamma\left(3+2 \kappa-\left(s_{3}+s_{4}+s_{5}+s_{6}\right)\right)} \nonumber\\
        &&  e^{\pi i\left(\frac{3}{2}+\kappa+2 \epsilon-2 s_{3}-2 s_{4}\right)}  \,,
    \end{eqnarray}
       where $i = 3,4,5,6$. However, the $B_{2c}^{(1)++, <}$ contour integral does not face this issue, since the pole should be included when we close the contour to the right, as shown in figure \ref{fig:pole} (d). 

    For $B_{2c}^{(1)++,>}$, the contour choices can be made as follows,
    \begin{equation}
        \epsilon = \frac{163}{44}, \ \mathcal{C}_2 = \frac{61}{48}, \  \mathcal{C}_3 = \frac{553}{384},\  \mathcal{C}_4 = \frac{131}{96},\     \mathcal{C}_5 = \frac{695}{384},\   \mathcal{C}_6 = \frac{191}{96} \ .
    \end{equation}
The sum of the residues then gives us that the leading order divergence is $-\frac{1}{1152 \sqrt{\pi} \epsilon^2 }$. Notice that the fictitious pole Eq.\eqref{eq,b2cppodot} gives a divergence $-\frac{1}{2880 \sqrt{\pi} \epsilon^2}$, which we subtract so that the divergence from $B_{2c}^{(1)++,>} $ is given by
\begin{equation}
   -\frac{1}{1152 \sqrt{\pi}\epsilon^2} -\left(- \frac{1}{2880 \sqrt{\pi} \epsilon^2}\right) =  -\frac{1}{1920} \frac{1}{\sqrt{\pi} \epsilon^2} \ .
\end{equation}
The contour choice for $B_{2c}^{(1) ++,<}$, which satisfies the condition $s_2 + s_5 + s_6 -\frac{3}{4} -\frac{3\epsilon}{2} < 0$, can be made as
\begin{equation}
    \epsilon = \frac{327}{128},\  \mathcal{C}_2 = \frac{643}{512}, \  \mathcal{C}_3 = \frac{253}{128},\  \mathcal{C}_4 = \frac{741}{512}, \    \mathcal{C}_5 = \frac{163}{128},\   \mathcal{C}_6 = \frac{987}{512} \ . 
\end{equation}
Evaluating the sum of the residues gives us $-\frac{1}{1152 \sqrt{\pi} \epsilon^2 }$, and summing both $B_{2c}^{++,<}$ and $B_{2c}^{++,>}$ then yields
\begin{equation}
    -\frac{1}{1920} \frac{1}{\sqrt{\pi} \epsilon^2}   -\frac{1}{1152} \frac{1}{\sqrt{\pi} \epsilon^2}   = -\frac{1}{720} \frac{1}{\sqrt{\pi} \epsilon^2}  \ .
\end{equation}
     Since $B_{2c}^{(1)--}$ is just the complex conjugation of $B_{2c}^{(1)++}$, we can safely double the above divergence and obtain that the leading order divergence for $B_{2c}^{(1)++} + B_{2c}^{(1)--}$ is $-\frac{1}{360 \sqrt{\pi}\epsilon^2}$. 
     
    \renewcommand{\em}{}
    \bibliographystyle{utphys}
    \addcontentsline{toc}{section}{References}
    \bibliography{dCFTdraft-v4}
    
    \end{document}